\documentclass[10pt]{article}

\usepackage[margin=1.0in]{geometry}
\usepackage{showlabels}
\usepackage{amsmath,amsfonts,amssymb,mathtools}
\usepackage{hyperref}
\usepackage{graphicx}
\usepackage{xcolor}
\usepackage{subfig}
\usepackage{tikz}

\numberwithin{equation}{section}

\DeclareMathAlphabet\mathbfcal{OMS}{cmsy}{b}{n}

\newcommand{\beq}{\begin{equation}}
\newcommand{\eeq}{\end{equation}}
\newcommand{\diag}{\text{diag}}
\newcommand{\tr}{\text{Tr}}
\newcommand{\Er}{{\mathbfcal{E}}}

\begin{document}

\begin{titlepage}

\begin{flushright}
IFUM-1091-FT
\end{flushright}

\vspace{1.7cm}

\begin{center}
\renewcommand{\thefootnote}{\fnsymbol{footnote}}
{\huge \bf Charged and rotating multi-black holes}
\vskip 7mm
{\huge \bf in an external gravitational field}
\vskip 31mm
{\large {Marco Astorino$^{a}$\footnote{marco.astorino@gmail.com} and Adriano Vigan\`o$^{a,b}$\footnote{adriano.vigano@unimi.it} }}\\

\renewcommand{\thefootnote}{\arabic{footnote}}
\setcounter{footnote}{0}
\vskip 10mm
{\small \textit{$^{a}$Istituto Nazionale di Fisica Nucleare (INFN), Sezione di Milano \\
Via Celoria 16, I-20133 Milano, Italy}\\
} \vspace{0.2 cm}
{\small \textit{$^{b}$Universit\`a degli Studi di Milano}} \\
{\small {\it Via Celoria 16, I-20133 Milano, Italy}\\
}
\end{center}
\vspace{3.2 cm}
\begin{center}
{\bf Abstract}
\end{center}
{We construct analytical and regular solutions in four-dimensional General Relativity which represent multi-black hole systems immersed in external gravitational field configurations. 
The external field background is composed by an infinite multipolar expansion, which allows to regularise the conical singularities of an array of collinear static black holes. 
A stationary rotating generalisation is achieved by adding independent angular momenta and NUT parameters to each source of the binary configuration. 
Moreover, a charged extension of the binary black hole system at equilibrium is generated. 
Finally, we show that the binary Majumdar--Papapetrou solution is consistently recovered in the vanishing external field limit. 
All of these solutions reach an equilibrium state due to the external gravitational field only, avoiding in this way the presence of any string or strut defect.
}

\end{titlepage}

\newpage

\tableofcontents

\newpage

\section{Introduction}

Multi-black holes solutions are intriguing astrophysical objects both from the theoretical and from the phenomenological point of view.
On the theoretical side, these solutions disclose the non-linear nature of General Relativity and represent an important playground in which testing the laws of black hole mechanics. 
On the experimental side, the recent remarkable observations of gravitational waves~\cite{Abbott:2016blz} heavily rely on the interactions between two black holes in a binary system:
thus an analytical description of such a spacetime is of utmost relevance for the interpretation of the measurements.

Of course one of the main obstacle in modelling a stationary multi-gravitational sources system is to provide a mechanism to balance the gravitational attraction of the bodies.
Otherwise the system naturally tends to collapse.
Usually the equilibrium is granted by the introduction of cosmological struts which prop up the gravitating bodies, but these one-dimensional objects must be constituted by matter that violates physically reasonable energy conditions.
Alternatively in some cases cosmic strings of infinite length can be used to support the gravitational collapse.
In both cases these objects are symptomatic for conical singularities which plague the spacetime.
Other objects, such as the Misner string, which occur in the presence of the gravimagnetic parameter, also known as NUT charge, are sometimes used to prevent the merging~\cite{Clement:2019clb,Clement:2020gjy}, but these bring in also harmful issues such as closed timelike curves. 
Instead our main objective is to remain as close to phenomenology as possible, therefore we will try to avoid physical structures which are outside realistic experimental or observational range.

We are interested in stationary and axisymmetric models basically for mathematical reasons:
the Einstein equations enjoy some integrability properties which can be exploited to generate highly non-trivial solutions without resolving the equation of motion.
To take advantage of the symmetries underling axisymmetric and stationary General Relativity we will mainly use the inverse scattering technique invented by Belinski and Zakharov~\cite{Belinsky:1971nt}, which will be briefly summarised in section~\ref{sec:ism}.
Also the coupling of the Einstein theory to the Maxwell electromagnetic field preserves the integrability of the axisymmetric and stationary system, therefore we will explore the possibility of including electromagnetic charge to some multi-black hole configuration, even though it seems quite difficult that these charged objects can exist spontaneously in Nature, because the distribution of matter in the Universe is neutral on large scale.

Basically the only regular multi-black holes configurations known so far were the extremal ones where the gravitational attraction is balanced by the electromagnetic force, such as in the Majumdar--Papapetrou solution~\cite{Majumdar:1947eu,Papapetrou} or the magnetised dihole~\cite{Emparan:1999au}\footnote{C-metrics embedded in an external magnetic field, found by Ernst~\cite{ErnstNodal}, can represent also a couple of black holes, when considering their maximal extension.
However these objects are two copies of the same black hole, always causally disconnected, and not interacting with each other, thus not really physically realistic.}.
Only very recently a new physical mechanism able to maintain the equilibrium was presented in the realm of vacuum General Relativity~\cite{Astorino:2021dju}.
It consists in the introduction of an external gravitational field endowed with a number of integration constants related to the field multipolar expansion.
In~\cite{Astorino:2021dju} a simple model was studied, which represents a system of two black holes immersed in a dipole-quadrupole external gravitational field.
Here we extend such solution in several ways:
firstly, we consider an arbitrary number of black holes immersed in an external field described by an infinite number of multipole momenta.
Then, in section~\ref{sec:rotating} and~\ref{sec:charged}, we build the rotating and charged generalisation of~\cite{Astorino:2021dju}.

External gravitational fields represent a natural setting for multi-black holes systems, as recent gravitational waves detection proceeding from the center of galaxies confirms.
In particular it has been shown in~\cite{deCastro:2011zz} that the multipolar gravitational field, we will deem here, can be produced by a distribution of matter such as thin disks or rings, typical shape of gravitational objects such as galaxies or nebulae.
Anyway the solutions considered in this article will be pure vacuum solution, without any energy-momentum tensor.
In principle a distribution of matter might be possibly considered very far away from the black holes.
In this sense our solutions can be interpreted as local models for binary or multi-black hole configurations.
In a certain sense the metrics presented here have to be considered as the gravitational analogous of the stationary black holes in Melvin magnetic universe.
In fact, also in this latter case, the solution is a pure electrovacuum solution with no definite sources for the electromagnetic external field, therefore their feasibility remains in the proximity domain of the black bodies.

Single black holes in external multipolar gravitational field have been pioneered by Doroshkevich, Zel'dovich and Novikov~\cite{NovikovZeldovich}, later studied by  Chandrasekhar~\cite{Chandrasekhar:1985kt}, Geroch and Hartle~\cite{Geroch:1982bv}.
These solutions are known in the literature as deformed black holes.
The novelty of our proposal is to take advantage of the external field to sustain the black bodies and prevent their collision.
From a mathematical point of view this means that the solution can be regularised from conical singularities that usually affect multi-black hole metrics.

The plan of the paper is the following:
we begin by reviewing the inverse scattering method, which is the leading method for the generation of our solutions, in section~\ref{sec:ism}.
Then, we describe the main properties of the background metric which produces the ``distortion'' on the black holes sources in section~\ref{sec:background}:
there we discuss both internal and external deformations from the point of view of the gravitational multipoles, even if in the following we make use of the external deformations only.
Section~~\ref{sec:array} is devoted to the array of collinear and static black holes immersed in the external field, while in sections~\ref{sec:rotating} and~\ref{sec:charged} we present the rotating and charged generalisations, respectively, of the binary system found in~\cite{Astorino:2021dju}.
Finally, we summarise our work and present some conclusions.

\section{Inverse scattering method}
\label{sec:ism}

We make use of the inverse scattering method~\cite{Belinsky:1971nt,Belinsky:1979mh,Belinski:2001ph} to superimpose black holes on top of the background spacetime containing an external gravitational field.
Thus, we begin by discussing the key features of the solution generation technique and subsequently by constructing the quantities apt to generate the desired spacetime.

The inverse scattering method relies on the integrability of the Einstein equations for the class of stationary and axisymmetric spacetimes, which can be described by the general metric in Weyl coordinates~\cite{Weyl1917}
\beq
\label{ism-seed}
{ds}^2 = f(\rho,z) ({d\rho}^2 + {dz}^2) + g_{ab}(\rho,z) {dx}^a {dx}^b \, ,
\eeq
where $a,b=0,1$ and $x^0=t$, $x^1=\phi$.
The spacetime~\eqref{ism-seed} possesses two commuting Killing vectors proportional to $\partial_t$ and $\partial_\phi$.
We assume that the coordinate $\rho$ is chosen such that $\det g = -\rho^2$.

The vacuum Einstein equations
$R_{\mu\nu} = 0$
can be equivalently written as
\begin{subequations}
\label{einstein}
\begin{align}
\label{eqUV}
U_{,\rho} + V_{,z} & = 0 \, , \\
\label{eqf1}
(\log f)_{,\rho} & = -\frac{1}{\rho} + \frac{1}{4\rho} \tr (U^2-V^2) \, , \\
\label{eqf2}
(\log f)_{,z} & = \frac{1}{2\rho} \tr (UV) \, ,
\end{align}
\end{subequations}
where $U=\rho g_{,\rho} g^{-1}$ and $V=\rho g_{,z} g^{-1}$ are two $2\times2$ matrices.
We see that, by solving equation~\eqref{eqUV} for $g$, one is able to find the function $f$ in quadratures from equations~\eqref{eqf1} and~\eqref{eqf2}.
Thus, the problem of solving the vacuum Einstein equations is reduced to the problem of finding the matrix $g$.

One can show that the integrability condition for the Einstein equations~\eqref{einstein} is equivalent to the linear equations
\beq
\label{schrodinger}
D_1\Psi = \frac{\rho V-\lambda U}{\lambda^2+\rho^2} \Psi \, , \qquad
D_2\Psi = \frac{\rho U+\lambda V}{\lambda^2+\rho^2} \Psi \, ,
\eeq
for the generating matrix $\Psi(\rho,z,\lambda)$,
where the commuting differential operators $D_1$ and $D_2$ are given by
\beq
D_1 \coloneqq \partial_z - \frac{2\lambda^2}{\lambda^2+\rho^2} \partial_\lambda \, , \quad
D_2 \coloneqq \partial_\rho + \frac{2\lambda\rho}{\lambda^2+\rho^2} \partial_\lambda \, ,
\eeq
and $\lambda$ is a complex spectral parameter.

In the inverse scattering method we prepare a seed solution $(g_0,f_0)$, and then find a generating matrix $\Psi_0$ which satisfies the linear eigenvalue equations~\eqref{schrodinger}.
Given such a $\Psi_0$, we introduce the functions
\begin{subequations}
\label{solitons}
\begin{align}
\mu_k(\rho,z) & = \sqrt{\rho^2 + (z-w_k)^2} - (z-w_k) \, , \\
\bar{\mu}_k(\rho,z) & = -\sqrt{\rho^2 + (z-w_k)^2} - (z-w_k) \, ,
\end{align}
\end{subequations}
where $w_k$ are arbitrary (complex) constants, called poles.
$\mu_k$ and $\bar{\mu}_k$ are called solitons and anti-solitons, respectively, and they satisfy
$\mu_k \bar{\mu}_k = -\rho^2$.
We will make us of the solitons only, in the following.

We associate a 2-dimensional vector (called BZ vector) to each (anti-)soliton
\beq
\label{bz}
m_a^{(k)} = m_{0 \, b}^{(k)} \bigl[\psi_0^{-1}(\mu_k,\rho,z)\bigr]_{ba} \, ,
\eeq
with arbitrary constants $m_{0 \, a}^{(k)}$.
Given the matrix
\beq
\Gamma_{kl} = \frac{m_a^{(k)} (g_0)_{ab} m_b^{(l)}}{\rho^2 + \mu_k \mu_l} \, ,
\eeq
a new metric is obtained by adding $2N$ solitons to $(g_0,f_0)$ as
\begin{subequations}
\label{metric-ph}
\begin{align}
\label{g-ph}
g_{ab} & = \pm \rho^{-2N} \Biggl(\prod_{k=1}^{2N} \mu_k\Biggr) \Biggl[(g_0)_{ab} - \sum_{k,l=1}^{2N} \frac{(\Gamma^{-1})_{kl} L_a^{(k)} L_b^{(l)}}{\mu_k\mu_l}\Biggr] \, , \\
\label{f-ph}
f & = 16 C_f f_0 \rho^{-(2N)^2/2} \Biggl( \prod_{k=1}^N \mu_k^{2N+1} \Biggr) \Biggl[ \prod_{k>l=1}^{2N} (\mu_k-\mu_l)^{-2} \Biggr] \det\Gamma \, ,
\end{align}
\end{subequations}
where
$L_a^{(k)} = m_c^{(k)} (g_0)_{ca}$ and $C_f$ is an arbitrary constant.
The new metric~\eqref{metric-ph} satisfies by construction the Einstein equations~\eqref{einstein} and is such that
$\det g=-\rho^2$.

\section{Background metric: internal and external multipoles}
\label{sec:background}

We discuss the general solution to the Einstein equations in vacuum, which contemplates both the internal deformations of the source and the contributions which come from matter far outside the source.
This general solution finds its roots in the pioneeristic work of Erez and Rosen~\cite{erez-rosen}, and it was lately discussed and expanded in~\cite{NovikovZeldovich,Geroch:1982bv,Chandrasekhar:1985kt} to include the deformations due to an external gravitational field.

The general solution for the Weyl metric
\beq
\label{weyl}
{ds}^2 = -e^{2\psi(\rho,z)} {dt}^2 + e^{-2\psi(\rho,z)} \bigl[ e^{2\gamma(\rho,z)} ({d\rho}^2 + {dz}^2) + \rho^2 {d\phi}^2 \bigr] \, ,
\eeq
is given, following the conventions in~\cite{breton-manko}, by
\begin{subequations}
\begin{align}
\psi & = \sum_{n=1}^{\infty} \biggl( \frac{a_n}{r^{n+1}} + b_n r^n \biggr) P_n \, , \\
\gamma & = \sum_{n,p=1}^\infty \biggl[
\frac{(n+1)(p+1) a_n a_p}{(n+p+2)r^{n+p+2}} (P_{n+1} P_{p+1} - P_n P_p)
+ \frac{np b_n b_p r^{n+p}}{n+p} (P_n P_p - P_{n-1} P_{p-1}) \biggr] \, ,
\end{align}
\end{subequations}
where $r\coloneqq\sqrt{\rho^2+z^2}$ defines the asymptotic radial coordinate and $P_n=P_n(z/r)$ is the $n$-th Legendre polynomial.
The real constants $a_n$ describe the deformations of the source, while the real parameters $b_n$ describe the external static gravitational field.

We observe that the ``internal'' part $a_n$, which is related to the deformations of the source, is asymptotically flat:
this seems to contradict Israel's theorem~\cite{Israel:1967wq}, which states that the only regular and static spacetime in vacuum is the Schwarzschild black hole.
Actually, the internal deformations lead to curvature singularities not covered by a horizon, in agreement with the theorem.
Because of this feature, in the following sections we will discard the internal contributions and focus on the external ones only.

On the converse, the ``external'' part $b_n$ is not asymptotically flat.
This is in agreement with the physical interpretation:
this part of the metric represents an external gravitational field generated by a distribution of matter located at infinity.
This interpretation parallels the Melvin spacetime one, and in fact there is no stress-energy tensor here to model the matter responsible of the field.
The asymptotia is not flat because at infinity there is the source matter and the spacetime is not empty.
Actually it can be shown that the curvature invariants, such as the Kretschmann scalar, can grow indefinitely at large distances\footnote{For the black hole model we are considering below, these large distances can be quantified in several orders of magnitude larger than the scale of the black holes.} in some directions.
In this sense, a spacetime containing such an external gravitational field should be considered local, in the sense that the description is meaningful in the neighborhood of the black holes that one embeds in this background. In this regard these metric are not different with respect to the usual single distorted black hole studied in the literature \cite{breton-manko}, \cite{Abdolrahimi:2015gea}.
To have a completely physical solution, one should match the black holes immersed in the external field with an appropriate distribution of matter (such as galaxies), which possibly asymptotes Minkowski spacetime at infinity.
A model for matter content consistent with the multipolar background expansion treated here and based on a thin ring distribution can be found in~\cite{deCastro:2011zz}\footnote{Analysis of the sectors where the scalar curvature invariants are larger, at least for the firsts order of the multipole expansion, suggests that a ring matter distribution is the more appropriate one for these backgrounds.}.

Accelerating generalisations of the multipolar gravitational background studied in this section naturally are endowed with a Killing horizon of Rindler type.
These accelerating horizon are typically located in between the black hole sources and the far region~\cite{ManyAcc}.

The constants $a_n$ and $b_n$ are related to the multipole momenta of the spacetime $\mathcal{Q}_n$.
The relativistic definition of the multipole momenta was given by Geroch~\cite{Geroch:1970cd}, and lately refined by Hansen~\cite{Hansen:1974zz}.
That definition applies for a stationary, axisymmetric and asymptotically flat spacetime:
the modified Ernst potential  associated to the spacetime is expanded at infinity, and the first coefficients of the expansion correspond to the multipole momenta~\cite{Fodor}.
Clearly, the internal deformations asymptote a flat spacetime, while the external ones do not.

In the following, we compute the multipole momenta associated to the above deformations, in order to clarify the intepretation of $a_n$ and $b_n$.
In particular, we calculate the internal multipoles contributions by means of the standard definition, and then we propose a new approach for determining the multipoles associated to external deformations.
We shall see that the proposed definition is analogous to the usual one and gives a consistent interpretation of the external field parameters.

\subsection{Internal multipoles}

Let us start with the internal deformations.
If we turn off the $b_n$, then we are left with
\beq
\psi_\text{int} = \sum_{n=1}^{\infty} \frac{a_n}{r^{n+1}} P_n \, .
\eeq
We define $\xi$ as a function of the Ernst potential (cf.~Appendix~\ref{App-Harrison-Kramer-Neugebauer})
\beq
\Er = \frac{1-\xi}{1+\xi} \,,
\eeq
that in our case reads
\beq
\xi_\text{int} \coloneqq \frac{1-e^{2\psi_\text{int}}}{1+e^{2\psi_\text{int}}} \, .
\eeq
We want to expand the above expression around infinity:
following~\cite{Fodor}, we bring infinity to a finite point by defining
$\bar{\rho}=\rho/(\rho^2+z^2)$, $\bar{z}=z/(\rho^2+z^2)$, and by conformally rescaling $\xi_\text{int}$:
\beq
\bar{\xi} = \frac{1}{\bar{\rho}^2 + \bar{z}^2} \, \xi_\text{int} \, .
\eeq
One can prove that $\bar{\xi}$ is uniquely determined by its value on the $z$-axis:
since infinity corresponds to $\bar{\rho}=\bar{z}=0$ in the new coordinates, then we expand $\bar{\xi}$ around $\bar{z}$ for $\rho=0$
\beq
\label{int-exp}
\bar{\xi}(\bar{\rho}=0) = \sum_{j=0}^\infty M_j \bar{z}^j \, ,
\eeq
where $M_j$ are the expansion coefficients.
The multipole momenta are completely determined by the coefficients $M_j$, and in particular
one can show (see~\cite{Fodor} and references therein) that the first four coefficients $M_0,\dotsc,M_3$ are exactly equal to the first four multipole momenta $\mathcal{Q}_0^\text{int},\dotsc,\mathcal{Q}_3^\text{int}$.
Thus, for the internal deformations, we find
\beq
\label{poles-int}
\mathcal{Q}_0^\text{int} = M_0 = 0 \, , \quad
\mathcal{Q}_1^\text{int} = M_1 = -a_1 \, , \quad
\mathcal{Q}_2^\text{int} = M_2 = -a_2 \, , \quad
\mathcal{Q}_3^\text{int} = M_3 = -a_3 \, .
\eeq
We see that there is a direct correspondence between the coefficients $a_n$ and the multipole momenta, at least at the first orders.
$\mathcal{Q}_0^\text{int}$ is the monopole term, which is zero since no source (e.g.~no black hole) is present.
$\mathcal{Q}_1^\text{int}$ is the dipole, $\mathcal{Q}_2^\text{int}$ is the quadrupole and $\mathcal{Q}_3^\text{int}$ is the octupole moment.
The subsequent multipole momenta can be still computed from the expansion~\eqref{int-exp} by means of a recursive algorithm, but now they will be non-trivial combinations of the coefficients $M_j$~\cite{Fodor}.
For instance, the 16-pole is given by
$\mathcal{Q}_4^\text{int} = M_4 - 1/7 M_0^2 M_2$.

\subsection{External multipoles}

The above construction can not work for the external deformations:
the asymptotia is not flat, and infinity is the place where the sources of the external field are thought to be, hence it does not make sense to expand there.
On the converse, it is meaningful to detect the effects of the deformation in the origin of the Weyl coordinates:
thus, by paralleling the Geroch--Hansen treatise, we propose to expand the modified Ernst potential in the \emph{origin} of the cylindrical coordinates.

Now we consider only the external deformations
\beq
\psi_\text{ext} = \sum_{n=1}^{\infty} b_n r^n P_n \, ,
\eeq
with modified Ernst potential
\beq
\xi_\text{ext} = \frac{1-e^{2\psi_\text{ext}}}{1+e^{2\psi_\text{ext}}} \, .
\eeq
According to the above discussion, we assume that $\xi_\text{ext}$ is  completely determined on the $z$-axis, so we expand around the origin for $\rho=0$
\beq
\label{ext-exp}
\xi_\text{ext}(\rho=0) = \sum_{j=0}^\infty N_j z^j \, ,
\eeq
where $N_j$ are the expansion coefficients.
We \emph{define} the first four multipole momenta $\mathcal{Q}_0^\text{ext},\dotsc,\mathcal{Q}_3^\text{ext}$ as the coefficients $N_0,\dotsc,N_3$, which are equal to
\beq
\mathcal{Q}_0^\text{ext} = N_0 = 0 \, , \quad
\mathcal{Q}_1^\text{ext} = N_1 = -b_1 \, , \quad
\mathcal{Q}_2^\text{ext} = N_2 = -b_2 \, , \quad
\mathcal{Q}_3^\text{ext} = N_3 = \frac{b_1^3}{3} - b_3 \, .
\eeq
Again, the monopole moment is zero because of the absence of a source.
We observe, contrary to~\eqref{poles-int}, that the octupole $\mathcal{Q}_3^\text{ext}$ is given by a non-trivial mixing of the constants $b_1$ and $b_3$.
This definition occurs only for the first momenta:
a definition which takes into account higher momenta can be achieved by generalising the procedure in~\cite{Fodor}.

For the time being, we content ourselves by proposing the following interpretation:
the coefficients $b_n$ are related to the multipole momenta generated by the external gravitational field, similarly to what happens for the internal deformations.
Then, the presence of the external gravitational field affects the momenta of a black hole source immersed in it.

\subsection{Seed for the inverse scattering construction}

We are interested in the external deformations only, hence we consider $a_n=0$ hereafter, and focus only on the contributions from $b_n$.
We express the external gravitational field metric in a form which is suitable for the inverse scattering procedure, i.e.
\begin{subequations}
\label{seed}
\begin{align}
g_0 & = \diag\Biggl[ -\exp\biggr(2\sum_{n=1}^{\infty} b_n r^n P_n \biggr) ,
\rho^2 \exp\biggl(-2\sum_{n=1}^{\infty} b_n r^n P_n \biggr) \Biggr] \, , \\
f_0 & = \exp\Biggl[
2 \sum_{n,p=1}^\infty \frac{np b_n b_p r^{n+p}}{n+p} \bigl(P_n P_p - P_{n-1} P_{p-1}\bigr)
- 2\sum_{n=1}^{\infty} b_n r^n P_n
\Biggr] \, ,
\end{align}
\end{subequations}
The parameters $b_n$ are related the multipole momenta of the external field, as explained above.
Metric~\eqref{seed} represents a generic static and axisymmetric gravitational field.

Since we want to add black holes on the background represented by the gravitational field, we take the metric~\eqref{seed} as a seed for the inverse scattering procedure.
Following the discussion in section~\ref{sec:ism}, we need the generating matrix $\Psi_0$, which serves as starting point to build the multi-black hole solution.
The function which satisfies equations~\eqref{schrodinger} is~\cite{LetelierBrasil}
\beq
\label{psi0}
\Psi_0(\rho,z,\lambda) =
\begin{pmatrix}
-e^{F(\rho,z,\lambda)} & 0 \\
0 & (\rho^2 -2\lambda\rho - \lambda^2) e^{-F(\rho,z,\lambda)}
\end{pmatrix}
\, ,
\eeq
where
\beq
F(\rho,z,\lambda) = 2 \sum_{n=1}^\infty b_n
\Biggl[ \sum_{l=0}^\infty \binom{n}{l} \biggl(\frac{-\rho^2}{2 \lambda}\biggr)^l \biggl(z + \frac{\lambda}{2}\biggr)^{n-l}
- \sum_{l=1}^n \sum_{k=0}^{[(n-l)/2]}
\frac{(-1)^{k+l}2^{-2k-l} n! \lambda^{-l}}{k!(k+l)!(n-2k-l)!}
\rho^{2(k+l)} z^{n-2k-l} \Biggr] \, .
\eeq
Now we can construct the BZ vectors~\eqref{bz}:
we parametrise
$m_0^{(k)}=\bigl(C_0^{(k)},C_1^{(k)}\bigr)$,
where $C_0^{(k)}$, $C_1^{(k)}$ are constants that will be eventually related to the physical parameters of the solution.
The BZ vectors are thus
\beq
m^{(k)} = \biggl( -C_0^{(k)} e^{-F(\rho,z,\mu_k)}, 
\frac{C_1^{(k)}}{\mu_k} e^{F(\rho,z,\mu_k)} \biggr) \, .
\eeq
Depending on the value of $C_0^{(k)}$ and $C_1^{(k)}$, the spacetime will be static or stationary, as we will see in the following.

\section{Array of collinear static black holes in an external gravitational field}
\label{sec:array}

We now proceed to the generalisation of the Israel--Khan solution~\cite{Israel1964}, which represents an array of collinear Schwarzschild black holes.
The Israel--Khan metric is plagued by the presence of conical singularities which can not be removed by a fine tuning of the physical parameters\footnote{Actually, in the limit of an infinite number of collinear black holes, the conical singularities disappear and the metric is regular. See~\cite{Myers:1986rx}.}.
On the converse, we will see that the external gravitational field will furnish the force necessary to achieve the complete equilibrium among the black holes.

Given the seed metric~\eqref{seed} and the generating matrix~\eqref{psi0}, we construct a new solution by adding $2N$ solitons with constants
\beq
C_0^{(k)} =
\begin{cases}
1 & k \text{ even} \\
0 & k \text{ odd}
\end{cases}
, \qquad
C_1^{(k)} =
\begin{cases}
0 & k \text{ even} \\
1 & k \text{ odd}
\end{cases}
\, .
\eeq
This choice guarantees a diagonal, and hence static, metric.
Each couple of solitons adds a black hole, then the addition of $2N$ solitons gives rise to a spacetime containing $N$ black holes, whose metric is
\begin{subequations}
\label{n-bh}
\begin{align}
g & = \diag\Biggl[
-\frac{\prod_{k=1}^N \mu_{2k-1}}{\prod_{l=1}^N \mu_{2l}} \exp\Biggl(2{\sum_{n=1}^{\infty} b_n r^n P_n}\Biggr),
\rho^2 \frac{\prod_{l=1}^N \mu_{2l}}{\prod_{k=1}^N \mu_{2k-1}} \exp\Biggl(-2{\sum_{n=1}^{\infty} b_n r^n P_n}\Biggr)
\Biggr] \, , \\
\label{n-bh-fph}
\begin{split}
f & = 16C_f f_0
\Biggl( \prod_{k=1}^N \mu_{2k}^{2N+1}
\mu_{2k-1}^{2N-1} \Biggr)
\Biggl( \prod_{k=1}^{2N} \frac{1}{\rho^2+\mu_k^2} \Biggr)
\Biggl( \prod_{k=1,l=1,3,\cdots}^{2N-1} \frac{1}{(\mu_k-\mu_{k+l})^2} \Biggr) \\
&\quad\times
\Biggl( \prod_{k=1,l=2,4,\cdots}^{2N-2} \frac{1}{(\rho^2+\mu_k\mu_{k+l})^2} \Biggr)
\exp\Biggl[ 2 \sum_{k=1}^{2N} (-1)^{k+1} F(\rho,z,\mu_k) \Biggr] \, .
\end{split}
\end{align}
\end{subequations}
Metric~\eqref{n-bh} is, by construction, a solution of the vacuum Einstein equations~\eqref{einstein},
and it represents a collection of $N$ Schwarzschild black holes, aligned along the $z$-axis, and immersed in the external gravitational field~\eqref{seed}.

We limit ourselves to the case of real poles $w_k$, since it represents the physically most relevant situation.
These constants are chosen with ordering
$w_1<w_2<\cdots<w_{2N-1}<w_{2N}$
and with parametrisation
\beq
w_1 = z_1 - m_1\, , \quad w_2 = z_1 + m_1\, , \quad \dotsc \quad
w_{2N-1} = z_N - m_N\, , \quad w_{2N} = z_N + m_N \, .
\eeq
The constants $m_k$ represent the black hole mass parameters, while $z_k$ are the black hole positions on the $z$-axis.

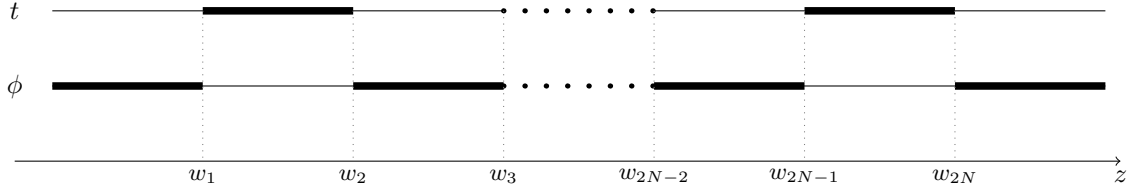
\begin{figure}
\centering
\begin{tikzpicture}
\draw[black,thin] (-10,1) -- (-4,1);
\draw[black,thin] (-10,2) -- (-4,2);
\draw[black,thin] (-2,1) -- (4,1);
\draw[black,thin] (-2,2) -- (4,2);

\draw[black,line width=1mm] (-10,1) -- (-8,1);
\draw[black,line width=1mm] (-8,2) -- (-6,2);
\draw[black,line width=1mm] (-6,1) -- (-4,1);

\draw[black,line width=1mm] (-2,1) -- (0,1);
\draw[black,line width=1mm] (0,2) -- (2,2);
\draw[black,line width=1mm] (2,1) -- (4,1);

\draw [line width=0.7mm,line cap=round,dash pattern=on 0pt off 4\pgflinewidth] (-4,2) -- (-2,2);
\draw [line width=0.7mm,line cap=round,dash pattern=on 0pt off 4\pgflinewidth] (-4,1) -- (-2,1);

\draw[gray,dotted] (-8,2) -- (-8,0);
\draw[gray,dotted] (-6,2) -- (-6,0);
\draw[gray,dotted] (-4,2) -- (-4,0);
\draw[gray,dotted] (-2,2) -- (-2,0);
\draw[gray,dotted] (0,2) -- (0,0);
\draw[gray,dotted] (2,2) -- (2,0);

\draw (-8,-0.2) node{{\small $w_1$}};
\draw (-6,-0.2) node{{\small $w_2$}};
\draw (-4,-0.2) node{{\small $w_3$}};
\draw (-2,-0.2) node{{\small $w_{2N-2}$}};
\draw (0,-0.2) node{{\small $w_{2N-1}$}};
\draw (2,-0.2) node{{\small $w_{2N}$}};
\draw (4.2,-0.2) node{$z$};

\draw (-10.5,2) node{$t$};
\draw (-10.5,1) node{$\phi$};

\draw[black,->] (-10.5,0) -- (4.2,0);
\end{tikzpicture}
\caption{{\small Rod diagram for the multi-black hole spacetime~\eqref{n-bh}.
The horizons correspond to the timelike rods (thick lines of the $t$ coordinate), while the conical singularities correspond to ``bolts'' where conical singularities can be avoided by imposing an appropriate periodicity on the angular coordinate.}}
\label{fig:rods}
\end{figure}

The black hole horizons correspond to the regions
$w_{2k-1}<z<w_{2k}$ ($k=1,\dotsc,N$), while the complementary regions are affected by the presence of conical singularities, as happens for the Israel--Khan solution (cf.~Fig.~\ref{fig:rods}).
Differently from that case, our solution can be regularised, as we will show in the next subsection.

It might be useful to write the metric~\eqref{n-bh} in the canonical Weyl form~\eqref{weyl}, in order to provide a comparison with the literature.
By writing down explicitly the solitons (according to~\eqref{solitons}) and the values for $w_k$, one finds the Weyl functions
\begin{align}
\psi & = \sum_{k=1}^N \psi_k + \sum_{n=1}^\infty b_n r^n P_n \, , \\
\gamma & = \sum_{k,l=1}^N \gamma_{kl} + \sum_{n,p=1}^\infty \frac{np b_n b_p r^{n+p}}{n+p} \bigl(P_n P_p - P_{n-1} P_{p-1}\bigr)
+ \sum_{k=1}^{2N} (-1)^{k+1} F(\rho,z,\mu_k) \, ,
\end{align}
where
\begin{align}
\psi_k & = \frac{1}{2} \sum_{k=1}^N \log\frac{R^+_k + R^-_k - 2m_k}{R^+_k + R^-_k + 2m_k} \, , \\
\gamma_{kl} & = \frac{1}{4} \log\frac{R^+_k R^-_l + (z-z_k-m_k)(z-z_l+m_l) + \rho^2}{R^-_k R^-_l + (z-z_k-m_k)(z-z_l-m_l) + \rho^2}
+ \frac{1}{4} \log\frac{R^+_k R^-_l + (z-z_k+m_k)(z-z_l-m_l) + \rho^2}{R^+_k R^+_l + (z-z_k+m_k)(z-z_l+m_l) + \rho^2} \, ,
\end{align}
and we defined
$R^\pm_k \coloneqq \sqrt{\rho^2 + (z-z_k\pm m_k)^2}$.
Actually, this form of the metric corresponds to the choice for $C_f$ made below (cf.~eq.~\eqref{n-cf}).
From this form, it is evident that the Israel--Khan solution~\cite{Israel1964} is recovered simply by putting $b_n=0$.

\subsection{Conical singularities and regularisation}
\label{sec:regularise}

The infinite multipole momenta $b_n$ allow to regularise the metric, i.e.~to remove all the conical singularities, by tuning their values.
More precisely, given $N$ black holes, there will be exactly $N+1$ conical singularities:
two cosmic strings, one rear the first black hole and one ahead the last black hole, and $N-1$ struts located between the $N$ black holes.
Hence, one will need at least $N+1$ parameters to fix the singularities.
We want to remove both the struts and the strings, since they are unphysical and lead to energy issues, as is shown in appendix~\ref{app:conic}.

The manifold exhibits angular defects when the ratio between the length and the radius of small circles around the $z$-axis is different from $2\pi$.
Working in Weyl coordinates, a small circle around the $z$-axis has radius $R=\sqrt{g_{zz}}\rho$ and length $L=2\pi\sqrt{g_{\phi\phi}}$~\cite{Astorino:2021dju}.
The regularity condition corresponds then to
$L/(2\pi R)\to 1$ as $\rho\to 0$.
It is easy to prove that, for the static and axisymmetric metrics of the class~\eqref{ism-seed}, the above condition is equivalent to $\mathcal{P}\equiv f g_{tt}\to 1$ as $\rho\to 0$.

We choose for convenience the gauge parameter $C_f$ as
\beq
\label{n-cf}
C_f = 2^{2(2N+1)} \Biggl[ \prod_{i=1}^N (w_{2i}-w_{2i-1})^2 \Biggr] \Biggl[
\prod_{k=1}^{N-1} \prod_{j=1}^{N-k}
(w_{2k-1} - w_{2k+2j})^2 (w_{2k} - w_{2k+2j-1})^2 \Biggr] \,  .
\eeq
The quantity $\mathcal{P}=f g_{tt}$ is equal to
\beq
\mathcal{P}_k =\Biggl[ \prod_{i=1}^{2k} \prod_{j=2k+1}^{2N}
(w_j-w_i)^{2\,(-1)^{i+j+1}} \Biggr]
\exp\Biggl[ 4\sum_{n=1}^\infty b_n \sum_{j=2k+1}^{2N} (-1)^{j+1} w_j^n \Biggl] \, ,
\eeq
between the $k$-th and $(k+1)$-th black holes (i.e.~$w_{2k}<z<w_{2k+1}$), for $1 \leq k < N$.
In the region $z<w_1$ we find
\beq \label{n-P0}
\mathcal{P}_0 =
\exp\Biggl[ 4\sum_{n=1}^\infty b_n \sum_{j=1}^{2N} (-1)^{j+1} w_j^n \Biggl] \, ,
\eeq
while for $z>w_{2N}$ we simply have
\beq
\mathcal{P}_N = 1 \, ,
\eeq
thanks to our choice of $C_f$.
These expressions are the natural generalisations of the conical singularities for the Israel--Khan metric~\cite{Gregory:2020mmi}.

The above expressions provide a system of equations
$\mathcal{P}_k=1$ for $0 \leq k < N$,
which can be solved, e.g., for the parameters
$b_1,\dotsc,b_N$, with the result of fixing all the conical singularities.
Hence the solution can be made completely regular outside the black hole horizons.

\begin{figure}[h]
\centering
\includegraphics[scale=0.41]{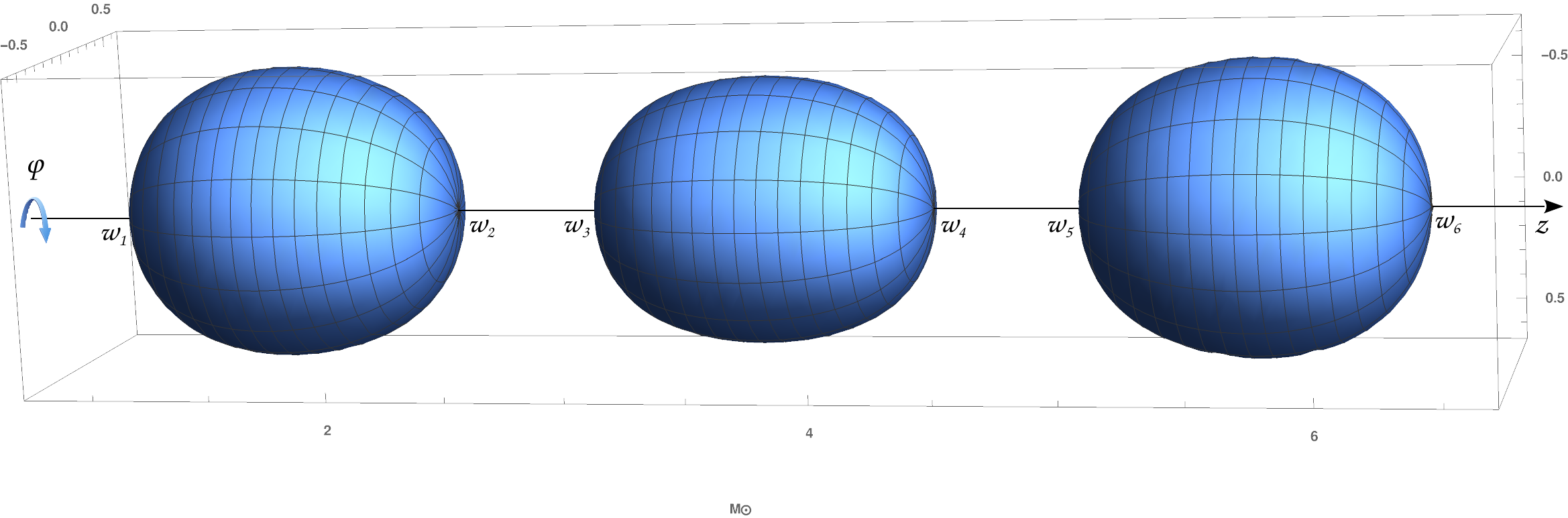}
\caption{\small Embedding diagram in $\mathbb{E}^3$ of the surfaces of three collinear black hole event horizons for the parametric values $w_1=1$, $w_2=12/5$, $w_3=3$, $w_4=22/5$, $w_5=5$, $w_6=32/5$ expressed in Solar mass units $M_\odot$.
The external gravitational background is endowed only with the first two terms of the multipolar expansion, the dipole and the quadrupole.
This picture shows the deformation of the horizons due to both the external gravitational field and the mutual interaction between the sources:
while all the sources have the same mass parameter (rod length), the central hole is more stretched because of the tidal forces provided by the external ones.
All the horizon surfaces are smooth because of the removal of any conical singularities thanks to the constraints on $b_n$ given by~\eqref{n-cf}-\eqref{n-P0}.
At most four parameters must be fixed in this specific triple source example:
$b_1=\frac{37}{56} \log{\frac{40000}{17901}}$,  $b_2=-\frac{5}{56} \log{\frac{40000}{17901}}$,
$b_n=0 \quad \forall n>2$, while $C_f=\frac{1606582813275738095616}{3814697265625}$.}
\label{fig:embedding3}
\end{figure}

As a pictorial example we show in a figure the geometry of the black holes array in the triple hole configuration.
As can be appreciated from Fig.~\ref{fig:embedding3}, the surfaces of the event horizons are regular because, thanks to the procedure explained above, the spacetime is everywhere devoid from conical defects.
At this purpose some integration constants, equal to the number of the spacelike rods of the rod-diagram, have to be fixed according to Eqs.~\eqref{n-cf}-\eqref{n-P0}.

\subsection{Smarr law}

We investigate the Smarr law for spacetime~\eqref{n-bh}:
to this end, we compute the total mass of the spacetime and the entropy and temperature of the black holes.

The mass is easily found by means of the Komar--Tomimatsu~\cite{Komar,Tomimatsu:1984pw} integral.
The result for the $k$-th black hole (i.e.~the black hole in the interval $w_{2k-1}<z<w_{2k}$) is
\beq
\label{n-mass}
M_k =
\alpha \int_{w_{2k-1}}^{w_{2k}} \rho g_{tt}^{-1} \partial_\rho g_{tt}
= \frac{\alpha}{2} (w_{2k}-w_{2k-1})
= \alpha m_k \, ,
\eeq
where $\alpha$ is a constant which takes into account the proper normalisation of the timelike Killing vector, generator of the horizon, $\xi=\alpha\partial_t$ associated to~\eqref{n-bh}.

The entropy of a black hole is related to the area as $S_k=\mathcal{A}_k/4$, hence
\beq
\label{n-entropy}
S_k =
\frac{1}{4} \lim_{\rho\to0} \int_0^{2\pi} d\phi \int_{w_{2k-1}}^{w_{2k}} dz \sqrt{f g_{\phi\phi}}
= \pi m_k  W
\exp\Biggl[ 2\sum_{n=1}^\infty b_n \sum_{j=2k}^{2N} (-1)^{j+1} w_j^n \Biggl] \, ,
\eeq
where
\beq
\log W = \lim_{\rho\to0} \log\sqrt{f g_{\phi\phi}} =
\log 2 + \sum_{i=1}^{2k-1} \sum_{j=2k}^{2N} (-1)^{i+j+1} \log|w_j-w_i| \, .
\eeq
The product $\sqrt{f g_{\phi\phi}}$ is independent of $z$ in the limit $\rho\to0$, and that was crucial in the derivation of~\eqref{n-entropy}.

Finally, the temperature is found via the Wick-rotated metric, and the result is
\beq
\label{n-temp}
T_k = \frac{\alpha}{2\pi} \lim_{\rho\to0} \rho^{-1} \sqrt{\frac{g_{tt}}{f}} =
\frac{\alpha}{2\pi} \lim_{\rho\to0} \frac{1}{\sqrt{f g_{\phi\phi}}} =
\frac{\alpha m_k}{2 S_k} \, .
\eeq
It is easily shown, by using~\eqref{n-mass},~\eqref{n-entropy} and~\eqref{n-temp}, that the Smarr law is satisfied:
\beq
\label{n-smarr}
\sum_{k=1}^N M_k = 2 \sum_{k=1}^N T_k S_k \, .
\eeq
We notice that the explicit value of $\alpha$ is not needed for~\eqref{n-smarr} to work, while it is relevant in the study of the thermodynamics~\cite{Astorino:2021dju}.

\subsection{Distorted Schwarzschild black hole}
\label{equilibrium-2-charged-BH}

The simplest non-trivial example we can consider for the complete external multipolar expansion from the general solution~\eqref{n-bh}, is clearly the single black hole configuration for $N=1$.
In that case the functions that appear in the Weyl static metric~\eqref{weyl} take the form
\begin{subequations}
\begin{align}
e^{2\psi} &=  -\frac{\mu_1}{\mu_2} \exp\biggr(2\sum_{n=1}^{\infty} b_n r^n P_n \biggr) \, , \\
e^{2\gamma} &=  \frac{16 \ C_f \ f_0 \ \mu_2^3 \ \mu_1 \  e^{2F(\rho,z,\mu_1)-2F(\rho,z,\mu_2)}}{(\rho^2+\mu_1^2)(\rho^2+\mu_2^2)(\mu_1-\mu_2)^2 } \, .
\end{align}
\end{subequations}
This spacetime represents a static black hole embedded in an external gravitational field.
The limit to the Schwarzschild metric is clear:
it is obtained just by switching off  all the multipoles $b_n=0 \quad \forall n$.
In order to recover the standard Schwarzschild metric in spherical coordinates the following transformation is needed
\beq
\rho = \sqrt{r(r-2m)}\sin\theta \, , \quad
z = z_1 + (r-m) \cos \theta \, .
\eeq
\begin{figure}[h!]%
\captionsetup[subfigure]{labelformat=empty}
\centering
\hspace{-0.2cm}
\subfloat[\centering $m=0.5$, $b_2=0.4$, $z_1=2$]{{\includegraphics[width=5.2cm]{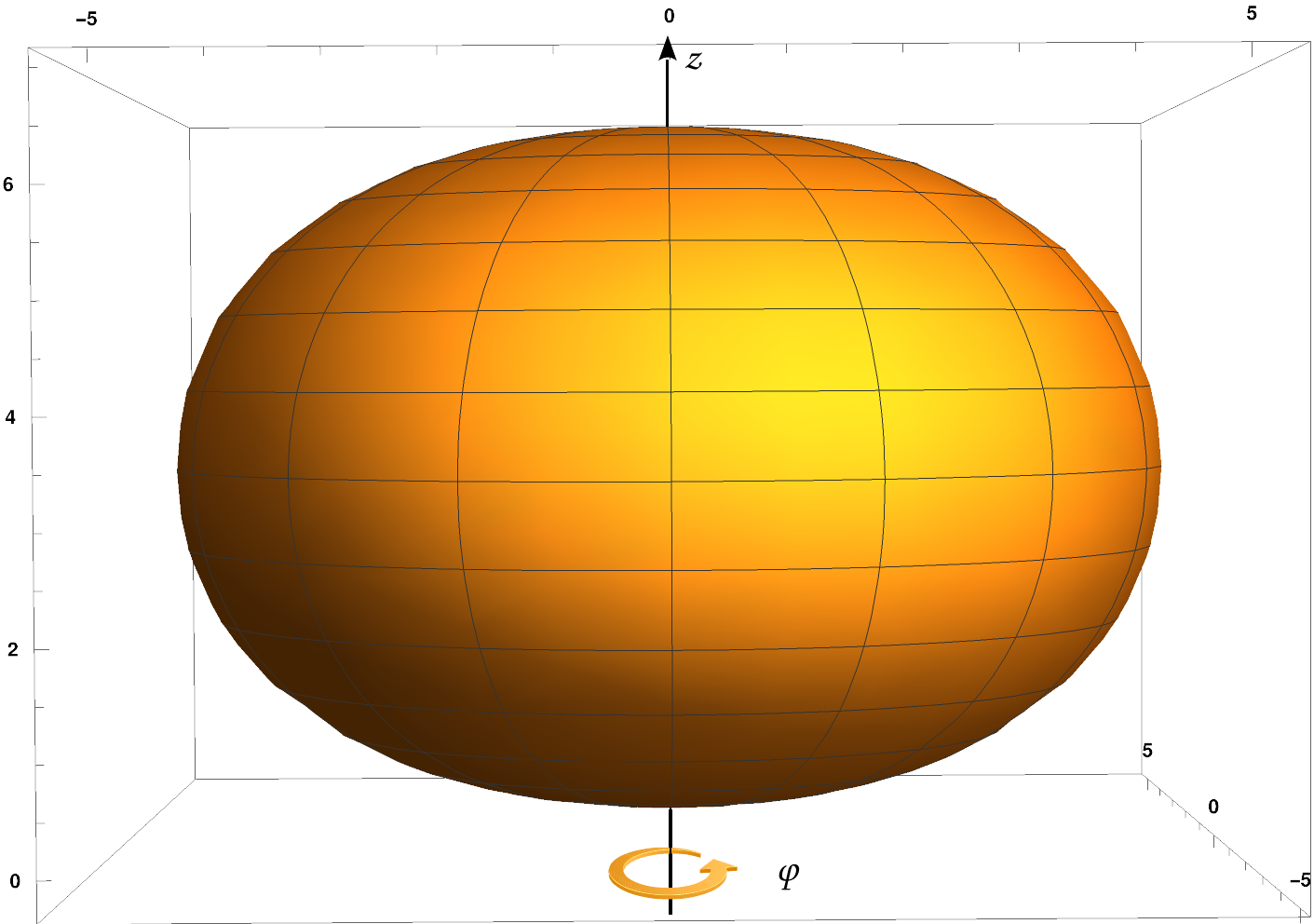}}}%
\subfloat[\centering $m=1$, $b_2=-0.1$, $z_1=2$]{{\hspace{-0.3cm} \includegraphics[width=7.5cm]{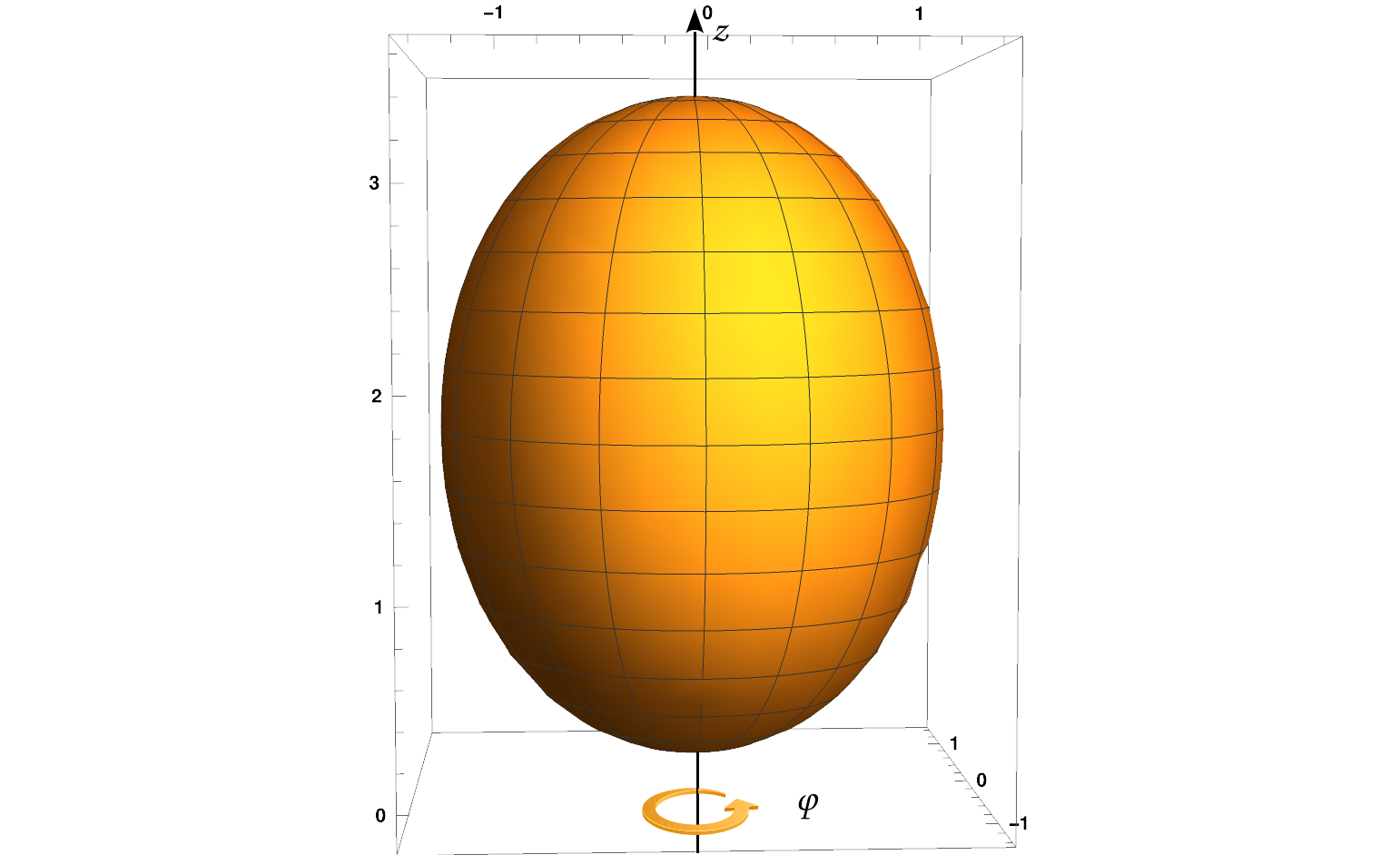}}}%
\subfloat[\hspace{-2.5cm} $m=1$, $b_2=-0.5$, $z_1=2$]{{\hspace{-2.5cm} \includegraphics[width=8.5cm]{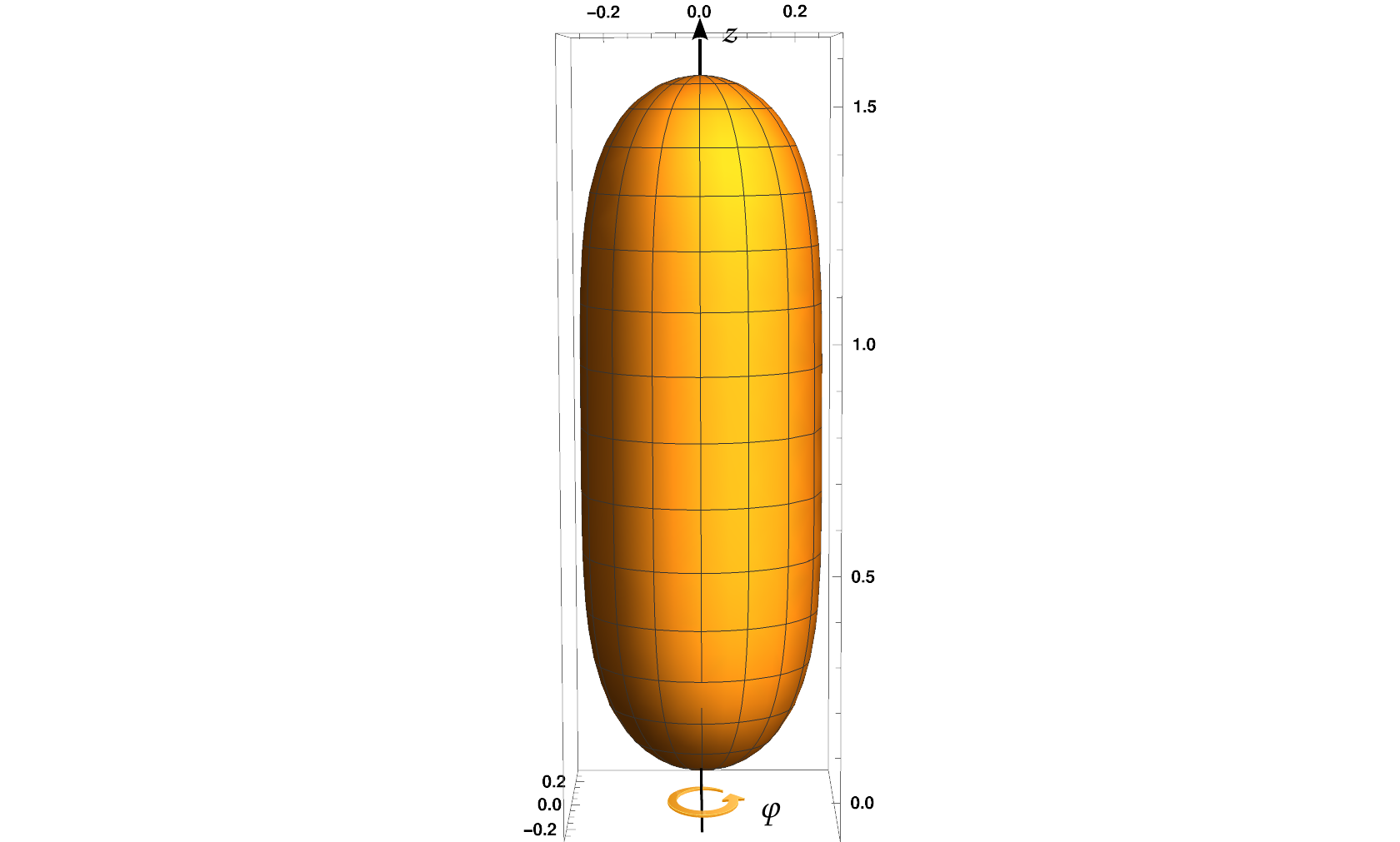}}}
\caption{\small Embeddings in euclidean three-dimensional space $\mathbb{E}^3$ of the event horizon of single black holes distorted by the external dipolar and quadrupolar gravitational field, for three different sets of physical parameters, expressed in Solar mass units $M_\odot$. $b_1$ and $C_f$ are fixed by the regularity constraint (\ref{costraints-sh1}).}%
\label{fig:bh-b1-b2}%
\end{figure}
A solution of this kind is not completely new, since it was already present in Chandrasekhar's book~\cite{Chandrasekhar:1985kt}, see also~\cite{breton-manko}.
However the form we are writing here is more general because, thanks to the extra parameter $z_1$, allows to place the black hole in any point of the $z$-axis.
In the absence of the external field the location of the black hole is irrelevant because the solution is symmetric under a finite shift of $z_1$.
But, when the external gravitational field is not null, a boost along the $z$-axis is significant, since the relative position of the black hole with respect to the external multipolar sources has some non-trivial effects on the geometry and on the physics of the black hole.

In particular the translation along the $z$-axis affects the event horizon shape.
In fact, as can be evaluated by computing the equatorial and polar circles around the event horizon, it is possible to understand as the horizon surface stretches or contracts depending on the position of the black hole and the values of the external parameters.
Some pictorial examples of the black hole horizon deformation for different external gravitational backgrounds are given in Fig.~\ref{fig:bh-b1-b2}.

\begin{figure}[h!]%
\centering 
\hspace{-3.5cm} 
\subfloat[\centering Black hole horizon with conical singularity]{{\includegraphics[width=13cm]{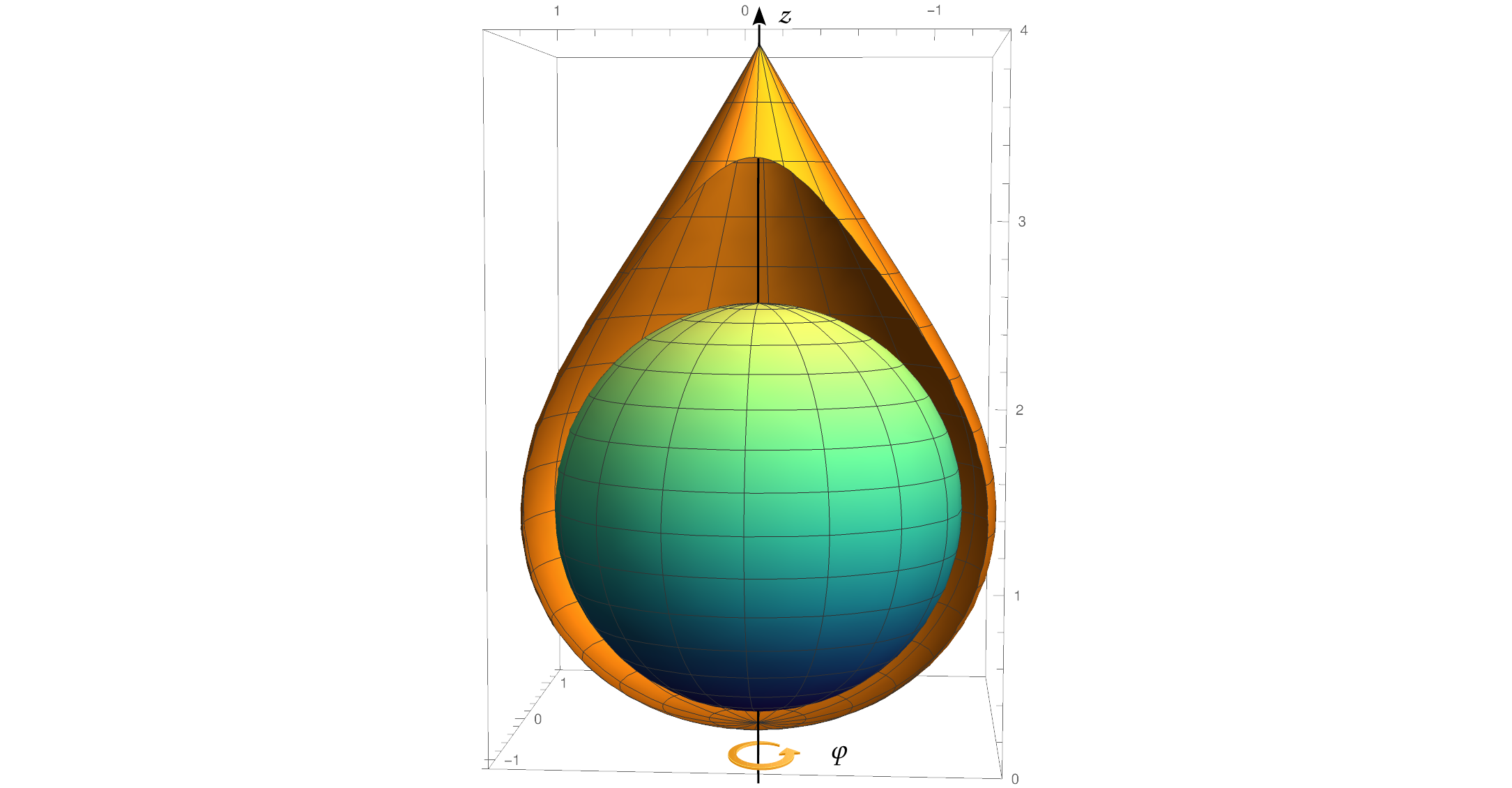}}}%
\hspace{-1.5cm}
\subfloat[\centering Black hole horizon regularised on both poles]{{\includegraphics[width=6.5cm]{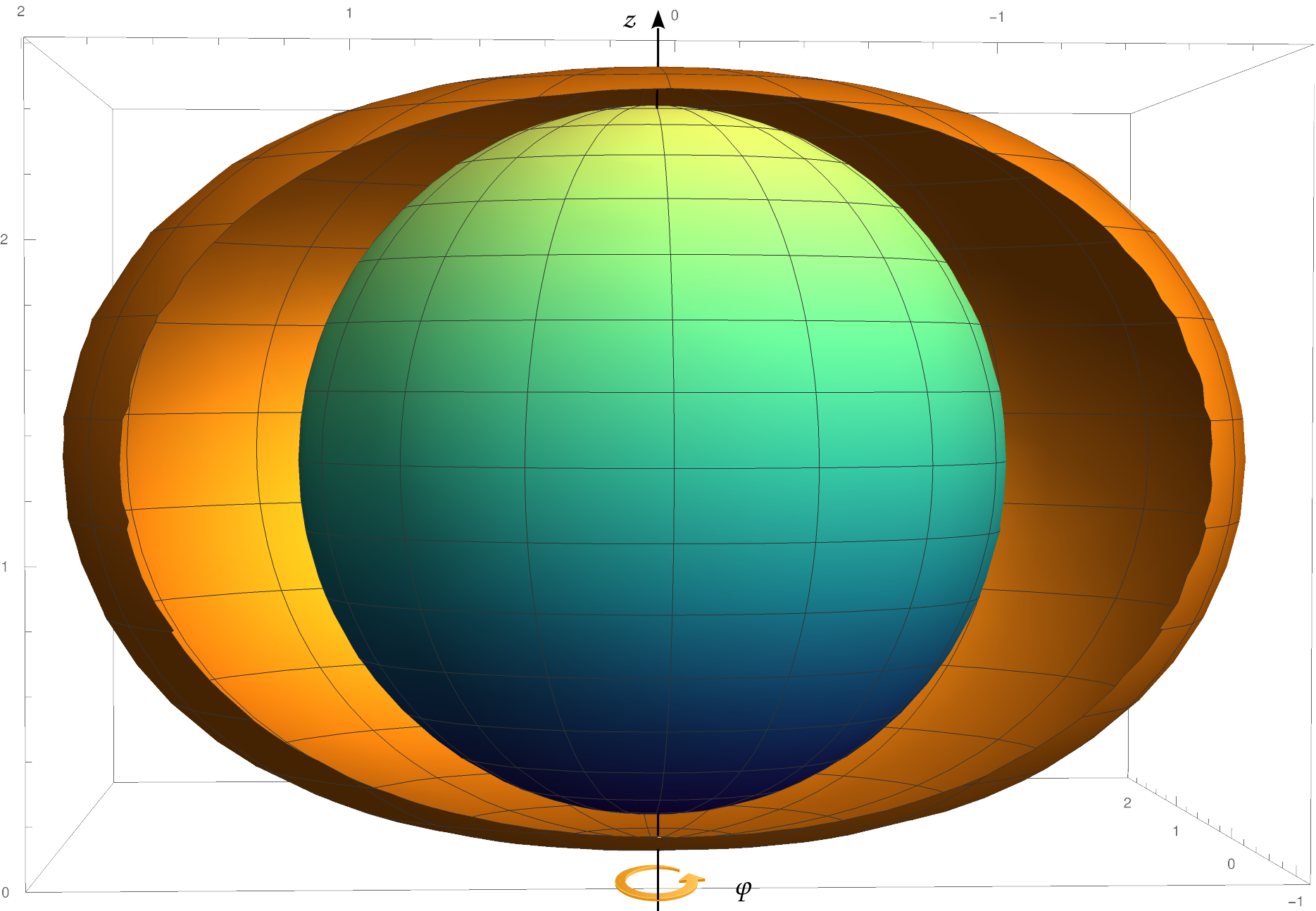} }}%
\caption{\small Two embedding in $\mathbb{E}^3$ of single static black hole horizons (yellow surfaces) immersed into a dipolar and quadrupolar external gravitational field, for $m = 0.6$, $z_1=-1.3$, $b_2=0.3$, expressed in Solar mass units $M_\odot$. A section is taken to appreciate the deformation with respect to the null external field case, drawn in green as a reference: the standard spherical Schwarzschild horizon, which is everywhere covered by the horizon swollen by the presence of the external gravitational field. The two black holes in external field differ only for the value of $b_1$, which  in Fig.~(b) is chosen according to Eq.~\eqref{costraints-sh1} to remove also the second conical singularity, while in Fig.~(a) $ b_1=0.5$.}%
\label{fig:con-sing}%
\end{figure}

Moreover, the equilibrium constraint which removes the conical defects can be loosen with respect to the one usually found in the literature~\cite{breton-manko}.
The imposition usually requested in the literature, i.e~$\sum_{n=1}^\infty b_{2n+1}=0$, is not fundamental when the black hole can be adjusted coherently with the external gravitational field.
In fact when $z_1 \neq 0$ the two regularising constraints we have to impose to avoid conical singularities become, as seen in section~\ref{sec:regularise},
\beq
\label{costraints-sh1}
C_f = \frac{(w_1-w_2)^2}{4} \, , \quad
\sum_{n=1}^\infty b_n \bigl( w_1^n - w_2^n \bigr) = 0 \, .
\eeq
When $b_n=0$ for $n>2$, we obtain a special subcase of the solution~\cite{Astorino:2021dju}, in the limit where one of the black hole vanishes or where the two horizon rods merge, remaining only with a single black hole configuration\footnote{This second limit is more easily obtained, in Weyl coordinates, when it is taken for the bi-dimensional Killing block of the metric $g_{ab}$~\eqref{g-ph}, and only subsequently the associate $f$ is generated according to the prescription~\eqref{f-ph}.}.
Fig.~\ref{fig:con-sing} in this section refer, for simplicity, to this truncated expansion of the external field.
A qualitatively analogous behaviour of the black hole horizon occurs in the full multipolar expansion. 

\section{Two Kerr--NUT black holes in an external gravitational field}
\label{sec:rotating}

We briefly discuss the rotating and NUTty generalisation of the binary black hole system immersed in an external gravitational field.
The generalisation to an array of rotating and NUT black holes is straightforward, at least conceptually.

We start again with the seed metric~\eqref{seed} and add four solitons (i.e.~two black holes).
However we make a different choice for the BZ constants:
we choose~\cite{Letelier:1998ft}
\begin{subequations}
\begin{align}
C_1^{(1)}C_0^{(2)} - C_0^{(1)}C_1^{(2)} & = \sigma_1 \, , \quad
C_1^{(1)}C_0^{(2)} + C_0^{(1)}C_1^{(2)} = -m_1 \, , \\
C_0^{(1)}C_0^{(2)} - C_1^{(1)}C_1^{(2)} & = n_1 \, , \quad
C_0^{(1)}C_0^{(2)} + C_1^{(1)}C_1^{(2)} = a_1 \, ,
\end{align}
\end{subequations}
and
\begin{subequations}
\begin{align}
C_1^{(3)}C_0^{(4)} - C_0^{(3)}C_1^{(4)} & = \sigma_2 \, , \quad
C_1^{(3)}C_0^{(4)} + C_0^{(3)}C_1^{(4)} = -m_2 \, , \\
C_0^{(3)}C_0^{(4)} - C_1^{(3)}C_1^{(4)} & = n_2 \, , \quad
C_0^{(3)}C_0^{(4)} + C_1^{(3)}C_1^{(4)} = a_2 \, .
\end{align}
\end{subequations}
Here $m_i$ are the mass parameters, $a_i$ are the angular momenta and $n_i$ are the NUT parameters.
We have also defined
$\sigma_i^2\equiv m_i^2 - a_i^2 + n_i^2$.
The poles $w_k$ are naturally defined as
\beq
w_1 = z_1 - \sigma_1 \, , \quad
w_2 = z_1 + \sigma_1 \, , \quad
w_3 = z_2 - \sigma_2 \, , \quad
w_4 = z_2 + \sigma_2 \, ,
\eeq
where $z_i$ are the positions of the black holes.

The resulting metric is computed by following the inverse scattering method of section~\ref{sec:ism}, but it is quite involved and we will not write it explicitly here.
The simplest form of the metric is achieved by using the bipolar coordinates
\beq
\begin{cases}
\rho = \sigma_1 \sqrt{x_1^2-1} \sqrt{1-y_1^2} \\
z = z_1 + \sigma_1 x_1 y_1
\end{cases}
, \quad
\begin{cases}
\rho = \sigma_2 \sqrt{x_2^2-1} \sqrt{1-y_2^2} \\
z = z_2 + \sigma_2 x_2 y_2
\end{cases}
\, .
\eeq
The metric regularisation from angular defects on the symmetry axis, can be pursuit as in the static case above, by tuning one physical parameter of the solution for each $N+1$ spacelike rod.

In the simplest case of a dipole-quadrupole configuration, such a binary system is characterised by ten parameters:
$\{ m_1,m_2,a_1,a_2,n_1,n_2,z_1,z_2,b_1,b_2 \}$.
Contrary to the standard double-Kerr case, in which there is no external field, the relevant physical quantity is not the distance between the holes $z_2-z_1$.
This happens because the external field does not act uniformly on the $z$-axis, hence the translation invariance along the axis is broken.
This means that the positions of the black holes are two independent parameters.

However, when the external field parameters are set to zero, $b_1=b_2=0$, one recovers the usual double-Kerr--NUT solution~\cite{Letelier:1998ft} (see~\cite{Manko:2018iyn} for a recent account on the equilibrium configurations).
Moreover, one can see that the positions $z_1$ and $z_2$ can be reabsorbed into a single parameter $l=z_2-z_1$.
Obviously, in such a limit the spacetime can not be regularised to give a physical solution, and the conical singularities can not be avoided.
The spacetime is then affected by the presence of struts or cosmic strings, unless one admits ``naked singularity-black hole'' or ``naked singularity-naked singularity'' configurations.

\section{Two Reissner--Nordstr\"om black holes in an external gravitational field}
\label{sec:charged}

A natural generalisation of the solutions presented in the previous sections contemplates the addition of the electric charge, thus we will construct a multi-Reissner--Nordstr\"om solution immersed in an external gravitational field.
Such a solution is the multi-black hole version of the metric presented in~\cite{breton-manko}.

\subsection{The charging transformation}

There are several procedures to extend a stationary and axisymmetric solution of general relativity to support a Maxwell electromagnetic field.
This methods allows, for instance, to generate the Reissner--Nordstr\"om spacetime from the Schwarzschild black hole solution, such as the Harrison~\cite{Harrison} or the Kramer--Neugebauer~\cite{KramerNeu} transformations.
We want to present here perhaps the simplest version of this charging transformation, which maps a given static and axisymmetric vacuum solution to another static and axisymmetric electrovacuum spacetime, typically adding monopole electric charge\footnote{These transformations combined with discrete symmetries of the spacetime can also act differently, for instance, they are best known for the ability to add an external electromagnetic background, of Melvin type, to the seed solution.}.

Let us consider the most general static and axisymmetric metric for Einstein--Maxwell theory, the Weyl metric~\eqref{weyl}.
Suppose the electromagnetic vector potential associated to this metric, compatible with the symmetries of this system, is null, i.e.~$A_\mu(\rho,z)=0$.
Then an electrically charged solution can be generated by the following transformation on the $\psi$ function of the metric
\beq
\label{hhat}
e^{2\psi} \to e^{2\hat{\psi}} = \frac{e^{2\psi}(1 - \zeta^2)^2}{(1-\zeta^2 e^{2\psi})^2}  \, ,
\eeq
which is be supported by an electric field given by
\beq
\label{Ahat}
\hat{A}_\mu =  \biggl( \frac{\zeta(e^{2\psi} - 1)}{1-\zeta^2 e^{2\psi}} , 0 ,0 ,0 \biggr) \, .
\eeq
The continuous parameter $\zeta$ can be considered real and it is related to the electric charge of the spacetime.
In appendix~\ref{App-Harrison-Kramer-Neugebauer} we derive this transformation from the Kramer--Neugebauer one and we show how the latter is contained in the Harrison transformation up to some gauge transformations.
There we also provide a simple example of application of the charging transformation~\eqref{hhat},~\eqref{Ahat}, where the electrically charged black hole solution of Reissner--Nordstr\"om is generated from the Schwarzschild metric.

\subsection{Generating two distorted Reissner--Nordstr\"om black holes}
\label{double-RN}

Now we want to charge a multi-Schwarzschild solution embedded in an external gravitational field.
To keep the model as simple as possible, without constraining the physical parameters of the black hole, we consider, as a seed, the double-black hole spacetime immersed in an external gravitational field possessing dipole and quadrupole moments only.
Note that this choice is done only for simplicity, but it could be chosen any external gravitational expansion endowed with multipoles of any order;
likewise the charging method allows one to deal easily with an arbitrary number of sources.
However we will act with the charging transformation~\eqref{hhat},~\eqref{Ahat} on the solution~\eqref{n-bh} for $N=2$ and with only the coefficients $b_1$ and $b_2$ different from zero.
The resulting seed metric, which coincides with the one described in~\cite{Astorino:2021dju}, is determined by the two functions of the static Weyl metric~\eqref{weyl}
\begin{subequations}
\begin{align}
e^{2\psi} & =  \frac{\mu_1\mu_3}{\mu_2\mu_4}
\exp\biggl[2b_1z + 2b_2\biggl(z^2 - \frac{\rho^2}{2}\biggr)\biggr] \, , \\
\label{gamma}
\begin{split}
e^{2\gamma} &=  \frac{ 16C_f \, e^{2\psi} \, \mu_1^3\mu_2^5\mu_3^3\mu_4^5}{W_{11}W_{22}W_{33}W_{44}W_{13}^2W_{24}^2Y_{12}Y_{14}Y_{23}Y_{34}}
\exp\biggl\{-b_1^2\rho^2 + \frac{b_2^2}{2} \bigl(\rho^2 - 8z^2\bigr)\rho^2 - 4b_1b_2 z \rho^2 \\
&\quad + 2 b_1 (-z + \mu_1 - \mu_2 + \mu_3 - \mu_4 )
+ b_2 \bigl[-2z^2 + \rho^2 + 4z (\mu_1 - \mu_2) + \mu_1^2 - \mu_2^2 \\
&\quad + (\mu_3 - \mu_4) (4z + \mu_3 + \mu_4) \bigr] \biggr\} \, ,
\end{split}
\end{align}
\end{subequations}
where
$W_{ij}=\rho^2+\mu_i\mu_j$ and $Y_{ij}=(\mu_i-\mu_j)^2$, while the solitons are defined as in~\eqref{solitons}.
The new solution\footnote{A Mathematica worksheet with this metric can be found among the files of this arXiv paper and on the web-page \href{https://sites.google.com/site/marcoastorino/papers/2105-02894}{https://sites.google.com/site/marcoastorino/papers/2105-02894}.} will maintain the same $\gamma$ as~\eqref{gamma} while transforming $\psi$ according to~\eqref{hhat}.

$\zeta$ is related to the electric charge of the black holes, therefore the transformed metric will represent a couple of charged black hole embedded in an external gravitational field.
In the limit of null external field $b_n=0$ and vanishing one of the two black hole masses ($m_2=0$ for instance), we exactly recover the Reissner--Nordstr\"om solution. Otherwise when the external gravitational field is not present the solution approach the equal mass double charged masses of \cite{Alekseev:2007gt}, \cite{Manko:2007hi}.

We are interested in solutions regular outside the event horizons, therefore we have to consider the quantity $\mathcal{P}\equiv f g_{tt}$ to avoid the possible conical singularities of the charged metric.
In fact, when $\mathcal{P}$ differs from $1$, it takes into account the deficit or excess angle along the three regions of the axial axis of symmetry outside the black holes event horizons, i.e.~for $\rho=0$ and $z \in (-\infty,w_1)$, $z \in (w_2,w_3)$ and $z \in (w_4,\infty)$.
The solution is made regular from line singularities by imposing the following three constraints on the metric parameters:
\begin{align}
\label{Cf}
C_f  &= 16 (w_1-w_2)^2(w_2-w_3)^2(w_1-w_4)^2(w_3-w_4)^2 \, , \\
\label{b1}
b_1  &=  \frac{w_1^2 - w_2^2 + w_3^2 - w_4^2}{2(w_1-w_2)(w_1+w_2-w_3-w_4)(w_3-w_4)} \log \biggl[\frac{(w_1-w_3)(w_2-w_4)}{(w_2-w_3)(w_1-w_4)} \biggr] \, , \\
\label{b2}
b_2  &= -\frac{w_1 - w_2 + w_3 - w_4}{2(w_1-w_2)(w_1+w_2-w_3-w_4)(w_3-w_4)}  \log \biggl[\frac{(w_1-w_3)(w_2-w_4)}{(w_2-w_3)(w_1-w_4)} \biggr] \, .
\end{align}
While this remains formally the same regularisation constraint of the uncharged case, the physical meaning of the parameters is different, as can be easily understood from the single source case treated in appendix~\ref{App-Harrison-Kramer-Neugebauer}.
In fact the two black hole horizons are located in $\rho=0$ and
$z\in(w_1,w_2)$, $z\in(w_3,w_4)$,
where
\beq \label{wi-charged}
w_1 = z_1 - \sigma_1 \, , \quad
w_2 = z_1 + \sigma_1 \, , \quad
w_3 = z_2 - \sigma_2 \, , \quad
w_4 = z_2 + \sigma_2 \, , \quad
\eeq
Henceforward we consider $C_f$ and $b_n$ fixed, as in Eqs.~\eqref{Cf}-\eqref{b2}, in order to assure the absence of conical singularities.

Note that the proper distance between the two event horizon surfaces converges:
\begin{equation}
\ell = \int_{w_2}^{w_3} dz\sqrt{g_{zz}} \Big|_{\rho=0}    < \infty \, .
\end{equation} 
It means that the balancing condition is non-trivial and can be realised for a finite separation between the sources. 

\subsection{Charges and Smarr law}

The electric charge of each black hole can be computed thanks to the Gauss law~\cite{Emparan:2001bb}
\beq\label{charge}
Q_i = - \frac{1}{4\pi} \int_0^{2\pi} d\phi \int_{w_{2i-1}}^{w_{2i}} dz \, \rho \, g_{tt}^{-1} \partial_\rho A_t \big|_{\rho=0} =
\frac{2 \zeta \sigma_i}{1-\zeta^2} \, .
\eeq 
Note that non-null results occur only in the regions which define the event horizon of the black holes, as expected.
Also note that, since the charging transformation is a one parameter transformation, it adds only an independent electric charge to the system.
Thus the free physical parameters of the solution are five: $z_i$, $\sigma_i$ and $\zeta$.
Hence the two black holes cannot vary independently their electric charge.
More general solutions involving independent electric charge parameters can be built, but with more refined generating techniques such as~\cite{Alekseev80} (or~\cite{Sibgatullin}).

The mass of the charged black holes can be defined by evaluating, on their respective event horizon, the following integral\footnote{The normalisation of the timelike killing vector here is considered unitary. The generic normalisation factor $\alpha$ for the mass is reintroduced in eqs. (\ref{smarr-i}), (\ref{Mtot-Qtot}).}
\beq\label{mass-charged}
M_i = \frac{1}{4} \int_{w_{2i-1}}^{w_{2i}} dz \bigl( \rho \, g_{tt}^{-1} \partial_\rho g_{tt} - 2 A_t \partial_\rho A_t \bigr) \big|_{\rho=0} =
\frac{1 - 2 A_0 \zeta + \zeta^2}{1-\zeta^2} \sigma_i \bigg|_{A_0=0} = \frac{1 + \zeta^2}{1-\zeta^2} \sigma_i \, .
\eeq
Considering the mass as a local quantity, i.e.~defined close to the horizon, we can fix the gauge degree of freedom in the electric potential as $A_0=0$.
Of course other gauge fixings can be pursued, for instance requiring that the electric potential vanish at large radial distances $\sqrt{\rho^2+z^2} \to \infty$.

From the masses and electric charges of the black holes, eqs.~\eqref{charge} and~\eqref{mass-charged}, we can deduce, for any $\zeta \neq 1$, the value of 
\begin{equation}
\label{sigma}
\sigma_i = \sqrt{M_i^2 - Q_i^2}  \, .
\end{equation}
As expected the the masses and electric charges are not all independent, but they can be expressed just in terms of the parameters $M_i$ and $\zeta$. In fact the electric charges can be written, thanks to eqs.~\eqref{charge} and~\eqref{mass-charged}, as
\beq
\label{mass-charge}
Q_i = \frac{2\zeta}{1+\zeta^2} M_i \, .
\eeq
The entropy for each black hole is taken as a quarter of the event horizon area
\beq
S_i = \frac{1}{4} \int_0^{2\pi} d\phi \int_{w_{2i-1}}^{w_{2i}} dz  \sqrt{g_{zz} g_{\varphi\varphi}}\ \big|_{\rho=0} \, ,
\eeq
which gives
\begin{align}
S_1  & = \pi \frac{(w_2-w_1)^2(w_4-w_1)}{(w_3-w_1)(1-\zeta^2)^2} e^{-2b_1(w_2-w_3+w_4)-2b_2(w_2^2-w_3^2+w_4^2)} \, , \\
S_2 & = \pi \frac{(w_4-w_3)^2(w_4-w_1)}{(w_4-w_2)(1-\zeta^2)^2} e^{-2 w_4 (b_1+b_2w_4)} \, .
\end{align}
The temperature of the event horizons, computed as in the previous section, can be written as
\beq
T_i = \frac{\alpha \sigma_i}{2 S_i} \, .
\eeq
From equation~\eqref{mass-charge} it is easy to see that when the masses of the two black hole coincides, $M_1=M_2$, also $Q_1=Q_2$.
Then it is possible to take $\zeta$ as in the single Reissner--Nordstr\"om case treated in appendix~\ref{App-Harrison-Kramer-Neugebauer}
\beq
\zeta = \frac{M_1 - \sqrt{M_1^2-Q_1^2}}{Q_1} \, .
\eeq
In this symmetric case it can be straightforwardly checked that both the temperature and the surface area of the two black holes coincide.
Anyway the thermal equilibrium can be reached also for more general sources configurations.
Note that the event horizons become extremal in the limit case for the charging transformation, $\zeta \to 1$, as in the single black hole case.

The Coulomb electric potential $\Theta$ evaluated on both the event horizons takes the same value 
\beq
\Theta = - \xi^\mu A_\mu \big|_{\rho=0} =  \alpha(\zeta-A_0) - \alpha\Theta_\infty  \, .
\eeq
We have now all the ingredients at our disposal to verify the Smarr law, both for the single element 
\beq \label{smarr-i}
\alpha M_i = 2 T_i S_i - \Theta Q_i \, , 
\eeq
and, thus, for the double black hole configuration
\beq
M = \sum_{i=1}^2 2 T_i S_i - \Theta Q \, ,
\eeq
where we defined 
\beq \label{Mtot-Qtot}
M = \alpha \sum_{i=1}^2 M_i \, , \quad
Q = \sum_{i=1}^2 Q_i \, .
\eeq
Since distorted black holes have a preferred interpretation as local systems, we primarily focused on local quantities, basically defined on the horizon.
Nevertheless the above results hold also in the case one considers the presence of the asymptotic Coulomb potential.
The gauge freedom encoded in $A_0$ can be used to put to zero the value of the potential at large distance:
when $A_0=\zeta^{\text{sign}(b_2)}$, then  $\Theta_\infty=0$.

The above results are valid for any $\alpha$, the  normalisation parameter of the Killing vector that generate the horizon $\xi=\alpha \partial_t$. Then, in this context, $\alpha$ can practically regarded as unitary.
However, in discussing the first law of black hole thermodynamics, it is necessary to select a particular value for $\alpha$, as described in~\cite{Astorino:2021dju}, for a local point of view based on the assumption that the observer are located close the hole and they have no access to infinity.

This charged black binary configuration is one of the few multi-black hole examples where it is concretely possible to test the second law of black holes thermodynamics, as done for the uncharged case~\cite{Astorino:2021dju} or for the Majumdar--Papapetrou black holes~\cite{Astorino:2019ljy}.
For instance, when the system is isolated, it is easy to verify that for two configurations, with the same energy and background field, the disjoint state is always less entropic than a collapsed state, which can be thought as the final state:
$S_{\odot\odot} < S_{\bigodot}$.
Anyway, for different boundary conditions this charged case has a rich phase transitions scenario, from adiabatic merging to black hole brimming.
However it is outside the scope of this work and will be studied elsewhere.

\subsection{Majumdar--Papapetrou limit}

Inspecting the values of the regularising parameters $b_n$ in eqs.~\eqref{b1},~\eqref{b2} we notice that there is a special case for which they vanish:
that happens for $w_1=w_2$ and $w_3=w_4$, which, according to eqs. (\ref{wi-charged}), (\ref{sigma}) and (\ref{mass-charge}) corresponds to extremality, $M_i=Q_i$, or $\zeta=1$\footnote{While the charging transformation is not well defined for $\zeta=1$, the metric present no mayor drawback in this limit, besides the usual carried by extremal horizons.}.
In that case the standard Minkowskian asymptotics is retrieved.

That is not surprising because the charged double black hole configuration presented in section~\ref{double-RN} naturally contains a particular subcase of the Majumdar--Papapetrou solution~\cite{Majumdar:1947eu,Papapetrou}, the one which describes two identical black holes located along the $z$-axis.
The black holes have to be identical because the charging transformation~\eqref{hhat},~\eqref{Ahat} adds only one independent charge and the Majumbdar--Papapetrou solution possesses only extremal horizons.
In fact the Majumdar--Papapetrou solution is the only configuration, within the double Reissner--Nordstr\"om system~\cite{Alekseev:2007gt,Manko:2007hi}, that can reach the equilibrium by balancing the gravitational attraction thanks to the electric repulsion of the sources, without requiring any hyper-extremal event horizon, and thus preserving its black hole interpretation.

Actually the limit to this version of the Majumdar--Papapetrou metric, describing a twin couple of (extremal) charged black holes, can be obtained easily, just considering the extremal limit of the solution~\eqref{hhat},~\eqref{gamma}, i.e.~$M_i=Q_i$.
It is given by the following simple form\footnote{The Majumdar--Papapetrou solution describing a binary black hole system in Weyl coordinate can be found in~\cite{Astorino:2019ljy}.}
\begin{align}
{d\hat{s}}^{2} &= -e^{2\hat{\psi}}  {dt}^2 + e^{-2\hat{\psi}} \bigl( {d\rho}^2 + {dz}^2 + \rho^2 {d\phi} \bigr) \, , \\
\hat{A}_t & = \biggl( 1 + \frac{M_1}{\sqrt{\rho^2+(z-z_1)^2}} + \frac{M_2}{\sqrt{\rho^2+(z-z_2)^2}} \biggr)^{-1} \, ,
\end{align}  
where
\begin{equation}
e^{2\hat{\psi}} = \hat{A}_t^2 \, . 
\end{equation}

\section{Summary and Conclusions}

In this article several generalisation of the binary black hole system at equilibrium in an external gravitational field are built, thanks to various solution generating techniques.
First of all the complete infinite multipolar expansion for the external field is considered as a background.
These multipoles allow us to regularise an arbitrary number of collinear static Schwarzschild black holes at equilibrium.
Therefore, thanks to the external gravitational field, we are able to remove all the conical singularities of the Israel--Kahn solution.
The physical quantities computed for this infinite array fulfil the Smarr relation.

We notice that in the single deformed Schwarzschild case our solution is more general with respect to similar ones studied in the literature.
This is because the one presented here has an extra physical parameter related with the position of the black hole with respect to the multipolar distribution.
This novel feature is fundamental in order to have a more general balance constraint, hence an enriched physical description.

Then, with the aid of the inverse scattering method, we show how to add both the angular momentum and the NUT charge to each element of the black hole configuration.
In this stationary case the metric describes a deformed multi-Kerr--NUT system.

Moreover we present, thanks to the Kramer--Neugebauer transformation, an electrically charged extension of the binary system at equilibrium.
It represents two Reissner--Nordstr\"om black holes held in balance by the external gravitational field.
Also in this case it is verified that the physical quantities of the double solution satisfy the Smarr relation.
It is shown how in the charged case the constraint that assure the absence of conical singularities can be fulfilled also without the need of the external gravitational field because it is sufficient the balance between the gravitational attraction and the electric repulsion between the two black holes to maintain the equilibrium.
Therefore this peculiar configuration has a standard flat Minkowskian asymptotics and can not be nothing else than the symmetric Majumdar--Papapetrou bi-hole, which is naturally contained into our charged binary system.

The presence of external gravitational fields such as the one produced by a distribution of matter around the black holes can be used to regularise the angular defects for a wide family of binary black hole systems.
In our picture there is no need for struts or cosmic strings which are of uncertain experimental plausibility and of strident theoretical soundness.
The multi-sources description proposed here is local, that is we believe have some phenomenological pertinence in the proximity of the black holes, at least till the galactic heavy matter that produce the necessary external gravitational multipolar expansion.

We believe that these new, well behaved, solutions of the Einstein equations can be of some astrophysical interest and also good reference for the Numerical Relativity community.
In fact, until now, this latter was the only to posses tools to model and describe multi-black hole systems and to interpret observations.

It would be interesting to study the stability of these metrics, even though, we expect that in the long term the final state might be a merging of the binary system.
Therefore these models might be mainly useful in some metastable phase of the eventual collision between the black holes.

Several other generalisations could be performed relatively easily thanks to generating techniques both in the realm of General Relativity and in related gravitational theories for which the integrable methods are still applicable, from Brans--Dicke to other scalar tensor theories.

\section*{Aknowledgments}

This work was supported in part by Conicyt--Beca Chile, in part by MIUR-PRIN contract 2017CC72MK003 and also by INFN.

\appendix

\section{Conical singularities and energy conditions}
\label{app:conic}

Conical singularities, beyond making the spacetime manifold ill-defined from a mathematical point of view, give also rise to energy issues.
In general, such singularities can be interpreted as strings or struts whose energy-momentum tensor has a $\delta$-like nature.
We show what are the physical issues that the conical singularities bring in when they are present in the spacetime.

Let us consider Minkowski spacetime with an wedge of angle $2\pi\alpha\prime$ artificially removed.
By defining $C=1-\alpha\prime$, we can write the metric as
\beq
{ds}^2 = -{dt}^2 + {dr}^2 + C^2 r^2 {d\varphi}^2 + {dz}^2 ,
\eeq
where $0\leq\varphi<2\pi$.
One can regard this spacetime as a field sourced by a cosmic string or a strut~\cite{linet1985static}, whose non-vanishing energy-momentum tensor components are
\beq
T^t_t = T^\varphi_\varphi = 2\pi \mu \, \delta(x,y) ,
\eeq
where
\beq
\mu = \frac{1-C}{4C} ,
\eeq
and $\delta(x,y)$ is the two-dimensional delta function depending on the coordinates orthogonal to the $z$-axis $(x,y)$.
$\mu$ is interpreted as the tension of the filament source.

This result can be generalised to a generic four-dimensional spacetime~\cite{Vickers:1987az}, for which one finds that the Einstein equations
$G_{\mu\nu} = 8\pi T_{\mu\nu}$
give rise to an energy-momentum tensor
\beq
\label{stress-tensor}
T^0_0 = T^3_3 = 2\pi \mu \, \delta(x,y) ,
\eeq
where now
\beq
\mu = \frac{2\pi - C}{4C} .
\eeq
$C$ represents again the angular excess/deficit for the azimuthal angle.

Let us consider, e.g., a two-black hole spacetime from~\eqref{n-bh} ($N=2$):
the result~\eqref{stress-tensor} clearly shows that for $\mu>0$ (positive tension), the source acts as a string that pull a black hole.
This is the behaviour of the conical singularities that one finds at $z<w_1$ and $z>w_4$
There are no negative-energy issues in this case, but the string extends to infinity and the $\delta$ function gives rise to a divergent energy-momentum tensor.

In the case of $\mu<0$ (negative tension), the conical singularity in $w_2<z<w_3$ acts as a strut which pushes apart the two black holes.
The energy density associated to the energy-momentum tensor is negative (i.e.~the strut is composed of anti-gravitational matter) and there is again a divergence due to the $\delta$ function.



\section{Harrison and Kramer--Neugebauer charging trasformations}
\label{App-Harrison-Kramer-Neugebauer}

Both the Harrison and the Kramer--Neugebauer~\cite{KramerNeu} transformations are two symmetries of the Ernst equations for the Einstein--Maxwell theory.
The Ernst equations are a couple of complex differential equations governing the axisymmetric and stationary  
electrovacuum of General Relativity~\cite{ErnstII}
\begin{align}
\label{ernst-ee} 
\bigl( \textsf{Re} \, \Er + | \mathbf{\Phi} |^2 \bigr) \nabla^2 \Er & = \bigl( \overrightarrow{\nabla} \Er + 2 \mathbf{\Phi}^* \overrightarrow{\nabla} \mathbf{\Phi} \bigr) \cdot \overrightarrow{\nabla} \Er   \, ,       \\
\label{ernst-em}
\bigl( \textsf{Re} \, \Er + | \mathbf{\Phi} |^2 \bigr) \nabla^2 \mathbf{\Phi} & = \bigl( \overrightarrow{\nabla} \Er + 2 \mathbf{\Phi}^* \overrightarrow{\nabla} \mathbf{\Phi} \bigr) \cdot \overrightarrow{\nabla} \mathbf{\Phi} \, .
\end{align}
They are equivalent\footnote{Up to a couple of real equations for $\gamma$ which remains decoupled from the system, and therefore can be always integrated in a second instance.} to the standard Einstein--Maxwell equations for fields possessing two commuting Killing vectors, such as for the spacetimes we are considering in this paper: ($\partial_t$, $\partial_\varphi$).
Equations~\eqref{ernst-ee},~\eqref{ernst-em} are expressed in terms of the complex Ernst potentials, defined as 
\beq
\label{def-Phi-Er}
\Er(\rho,z) \coloneqq h - |\mathbf{\Phi}|^2 + i \chi \, , \quad
\mathbf{\Phi}(\rho,z) \coloneqq A_t + i \tilde{A}_\varphi \, ,
\eeq
where $\tilde{A}_\varphi(\rho,z)$ and $\chi(\rho,z)$ are obtained from\footnote{The differential operators here are intended in flat three-dimensional Euclidean space in cylindrical coordinates ($\rho,z,\varphi$).
For a brief review of the Ernst formalism and symmetries see~\cite{Astorino:2019ljy}}
\begin{align}
\label{A-tilde}
\overrightarrow{\nabla} \tilde{A}_\varphi & \coloneqq - h \rho^{-1} \overrightarrow{e}_\varphi \times (\overrightarrow{\nabla} A_\varphi - \omega  \overrightarrow{\nabla} A_t ) \, , \\
\label{h} \overrightarrow{\nabla} \chi & \coloneqq - h^2 \rho^{-1} \overrightarrow{e}_\varphi \times \overrightarrow{\nabla} \omega - 2 \textsf{Im} (\mathbf{\Phi}^*\overrightarrow{\nabla} \mathbf{\Phi} )  \, .
\end{align}
It is a simple matter to verify that the complex Ernst equations~\eqref{ernst-ee},~\eqref{ernst-em} enjoy an SU(2,1) Lie-point symmetry group spanned by the finite transformations
\begin{subequations}
\label{su21-transf}
\begin{align}
\label{su21-i}
\Er \longrightarrow \hat{\Er} = \lambda \lambda^* \Er &\ ,  \quad \mathbf{\Phi} \longrightarrow  \hat{\mathbf{\Phi}} = \lambda \mathbf{\Phi} \, , \tag{I} \\
\label{su21-ii}
\Er \longrightarrow \hat{\Er} = \Er + i \ b & \ , \quad \mathbf{\Phi} \longrightarrow  \hat{\mathbf{\Phi}} = \mathbf{\Phi} \, , \tag{II} \\
\label{su21-iii}
\Er \longrightarrow \hat{\Er} = \frac{\Er}{1+ic\Er} & \ , \quad  \mathbf{\Phi} \longrightarrow  \hat{\mathbf{\Phi}} = \frac{\mathbf{\Phi}}{1+ic\Er} \, , \tag{III} \\
\label{su21-iv}
\Er \longrightarrow \hat{\Er} = \Er - 2\beta^*\mathbf{\Phi} - \beta\beta^* & \ , \quad \mathbf{\Phi} \longrightarrow  \hat{\mathbf{\Phi}} = \mathbf{\Phi} + \beta  \, , \tag{IV} \\
\label{su21-v}
\Er \longrightarrow \hat{\Er} = \frac{\Er}{1-2\alpha^*\mathbf{\Phi}-\alpha\alpha^*\Er} & \ , \quad \mathbf{\Phi} \longrightarrow  \hat{\mathbf{\Phi}} = \frac{\mathbf{\Phi}+\alpha\Er}{1-2\alpha^*\mathbf{\Phi}-\alpha\alpha^*\Er} \, . \tag{V}
\end{align}
\end{subequations}
Greek letters ($\lambda$, $\beta$, $\alpha$) represent continuous complex parameters, while latin letters, such as ($b$, $c$), label real ones.

Some of these transformations, such as~\eqref{su21-i},~\eqref{su21-ii} and~\eqref{su21-iv} are pure gauge transformations, hence they could be reabsorbed by a diffeomorphism.
The Harrison transformation is~\eqref{su21-v}, while the Kramer--Neugebauer, as defined in~\cite{Manko_1992} to charge the Kerr metric embedded in an external gravitational field is
\beq
\label{KN}
\text{(KN)} \qquad  \Er \longrightarrow \Er' = \frac{\Er-\zeta^2}{1-\zeta^2 \Er} \,  , \quad \mathbf{\Phi} \longrightarrow  \hat{\mathbf{\Phi}} = \frac{\zeta (\Er-1)}{1-\zeta^2 \Er}\ \, .
\eeq
The latter transformation reduces to the one in~\eqref{hhat},~\eqref{Ahat} for static and uncharged seeds.
Since both the Kramer--Neugebauer~\eqref{KN} and the Harrison transformation~\eqref{su21-v}, have the same physical effects (they are know to add an electric monopole to an uncharged seed), we have the suspect that they are basically the same transformation, up to gauge transformations.
In fact it can be shown that the subsequent composition of transformations~\eqref{su21-i},~\eqref{su21-iv} and~\eqref{su21-v} to a Ernst seed $(\Er_0,\Phi_0)$ gives
\beq
\text{(V)} \circ \text{(IV)} \circ \text{(I)}  \circ
\begin{pmatrix}
\Er_0  \\
\Phi_0
\end{pmatrix}
=
\begin{pmatrix}
\frac{\lambda \lambda^* \Er_0 - \beta^*(\beta + 2 \lambda \Phi_0)}{1 + \alpha^2(-2 \beta + \alpha^* \beta \beta^* - \alpha \lambda \lambda^* \Er + 2\lambda (\alpha \beta^*-1) \Phi_0} \\
\frac{\beta - \alpha \beta \beta^* + \lambda\lambda^* \alpha \Er_0 + \Phi_0 -2\lambda\alpha\beta^* \Phi_0}{1 + \alpha^2(-2 \beta + \alpha^* \beta \beta^* - \alpha \lambda \lambda^* \Er + 2\lambda (\alpha \beta^*-1) \Phi_0}                         
\end{pmatrix}
\, .
\eeq
Then considering a null electromagnetic Ernst potential, $\Phi_0=0$, the imaginary part of the parameters $\alpha$, $\beta$, $\lambda$ zero and choosing the specific values
\beq
\lambda = \frac{1}{1-\zeta^2} \, ,
\quad \alpha = \zeta \, ,
\quad \beta = - \frac{\zeta}{1-\zeta^2} \, ,
\eeq
we exactly recover the (KN) transformation~\eqref{KN}.
In case of static metrics the latter further simplifies to~\eqref{hhat},~\eqref{Ahat}.
Therefore $(KN)$ and $(V)$ are basically equivalent, up to gauge transformations, so they might be called collectively Harrison-Kramer-Neugebauer transformation.  

As an explicit example we show the efficacy of the charging transformation~\eqref{hhat},~\eqref{Ahat} on an asymptotically flat, static and discharged metric.
For instance acting on the Schwarzschild metric, we produce the Reissner--Nordstr\"om black hole.

For simplicity we take the seed in spherical symmetric coordinates
\beq
{ds}^2 = -\biggl(1- \frac{2m}{r} \biggr) {dt}^2 + \frac{{dr}^2}{1-\frac{2m}{r}} + r^2 {d\theta}^2 + r^2 \sin^2 \theta {d\varphi}^2 \, ,
\eeq
from which we can easily read the seed
\beq
e^{2\psi} = 1-\frac{2m}{r}  \, , \quad A_t = 0 \, .
\eeq 
After the charging transformation~\eqref{hhat},~\eqref{Ahat} we get the new solution
\beq
e^{2\hat{\psi}} = \frac{r(r-2m)(1-\zeta^2)^2}{[r+(2m-r)\zeta^2]^2} \, , \quad \hat{A}_t = -\frac{2m\zeta}{r+(2m-r)\zeta^2} \, .
\eeq
A shift of the radial coordinate 
\beq
r \to \hat{r} - M + \sqrt{M^2-q^2} \, ,
\eeq
and a rescaling of the parameters
\beq
\label{zeta}
\zeta \to \frac{M - \sqrt{M^2-q^2}}{q} \quad , \qquad  m \to \sqrt{M^2-q^2} \, ,
\eeq
is sufficient to recognise the standard Reissner--Norstr\"om spacetime
\beq
{d\hat{s}}^2 = - \biggl(1-\frac{2 M}{\hat{r}}+\frac{q^2}{\hat{r}^2} \biggr){dt}^2 + \frac{{d\hat{r}}^2}{1-\frac{2M}{\hat{r}}+\frac{q^2}{\hat{r}^2}}+ \hat{r}^2 {d\theta}^2 + \hat{r}^2 \sin^2 \theta {d\varphi}^2 \, ,
\eeq
\beq
       \hat{A}_\mu = \left( - \frac{q}{\hat{r}},0,0,0 \right) \ .
\eeq
Note that this procedure, even though it always produces a non-equivalent (to the seed) solution, it might act not so predictably on an non-asymptotically flat solution, even if static.

\bibliographystyle{unsrturl}
\bibliography{RefLong.bib}

\end{document}